# THz-Driven Quantum Ionic Magnetism in a Quantum Paraelectric $SrTiO_3$

In Hyeok Choi[1,†], Sergei Urazhdin[2], Man Tou Wong[1], Zi-Jie Liu[1], and Keith A. Nelson[1,*]

[1]*Department of Chemistry, Massachusetts Institute of Technology, Cambridge, Massachusetts 02139, United states*
[2]*Department of Physics, Emory University, Atlanta, Georgia 30322, United states*

[†] First author

[*]Corresponding authors: kanelson@mit.edu

**Abstract**

Magnetic moments carried by rotating ionic motion in crystals are becoming recognized as an important contribution to magnetism, angular momentum transport, and optical activity. However, efficient approaches to their dynamical control are lacking. Here, we report THz-driven generation and optical detection of quantum ionic magnetism in quantum paraelectric $SrTiO_3$. We observe an oscillatory ionic magnetization without a corresponding oscillatory polarization, contradicting from the classical relation $\boldsymbol{M} \propto \boldsymbol{P} \times \dot{\boldsymbol{P}}$, while its suppression above the quantum paraelectric regime points to a quantum ionic origin. Analysis shows that this effect results from the beating between quantum ionic eigenstates whose degeneracy is lifted due to the directional symmetry breaking by the THz pulse. The presented approach provides a pathway for the ultrafast control of ionic magnetization in quantum materials for spintronic and thermotronic applications.

Angular momentum carried by circularly polarized light can be transferred to charge, spin and lattice systems in quantum materials via light-matter interaction. In accordance with the angular momentum conservation laws, this enables spin-selective optical measurements[1-9] and opto-spintronic applications[10-12], providing an additional tuning knob to control the degrees of freedom. Recently, it was shown that circularly polarized light can induce lattice angular momentum in an ionic crystal by generating axial phonons — circulating collective ion motion breaking time-reversal symmetry[6,12-16]. Similar axial phonon excitation through stimulated Raman scattering is possible using optical pulse pairs with different polarizations[17], a strategy also used to initiate unidirectional rotational motion in molecular gases[18].

Axial phonons can not only transport angular momentum but also transfer it to other degrees of freedom such as electron spins, enabling functionalities not limited by electrical conduction[19-21]. Early on, the magnetic moments of axial phonons were assumed to be negligible compared to their electronic counterparts, since they are scaled by the ratio of electron to nuclear mass. However, recent findings suggest that strong coupling to the electronic degrees of freedom can make them comparable[9,12-14,21-27], opening new pathways for axial-phonon-driven spintronic and thermotronic functionalities including ultrafast switching of magnetization in ferromagnetic layers[9,12], thermal-gradient-induced spin current generation[28,29], and chirality-selective thermal transport[30,31].

Axial ionic motion was conventionally assumed to be classical. However, if large ionic moments arise from the coupling to electrons, quantum-mechanical effects must be important since the electronic spin and orbital momenta are generally of quantum origin. The role of quantum mechanisms can be revealed by studies of quantum paraelectrics (QPEs) – paraelectric (PE) materials at the verge of ferroelectric (FE) transition. Their FE ordering at cryogenic temperatures is prevented by quantum ionic fluctuations, resulting in a non-classical ionic state[32-34]. Recently, it has been reported that the superposed quantum states of ionic positions can be disrupted by linearly polarized THz excitation on an ultrafast timescale, resulting in the generation of transient macroscopic polarization. This optical sensitivity of QPE states suggests the possibility of ultrafast control of magnetization arising from quantum ionic motion under circularly polarized THz excitation.

Here we report the observation of quantum ionic magnetism in archetypal QPE $SrTiO_3$ (STO) that can be unambiguously identified with ionic dynamics. At 105 K, STO undergoes an antiferrodistortive (AFD) transition from the high-temperature cubic PE phase to the centrosymmetric tetragonal 4/*mmm* phase, followed by the emergence of the QPE state below 40 K[35-37]. Under resonant elliptically polarized THz excitation with the transverse optical (TO) mode, we observed temperature-dependent sub-THz magnetic oscillations below 60 K, where quantum fluctuations become increasingly dominant. While a previous study reported large magnetic moments in the PE phase of STO that appear only during TO phonon oscillations[14], we found that the magnetic oscillations persist far beyond the relaxation time of TO phonon oscillations, indicating a quantum-mechanical origin. Our results highlight the significance of ionic magnetism and point to the importance of quantum-mechanical mechanisms.

To monitor ultrafast ionic magnetization in STO, we employed a THz-pump second-harmonic generation circular dichroism (SHG-CD) technique, as illustrated in Fig. 1a. Ionic dynamics were induced by the single-cycle elliptically polarized THz pulses, with the center of ellipticity frequency $f_0$ = 0.7 THz close to the TO phonon frequency $f_{TO}$ = 0.5 THz of STO in the QPE state (Section S1, Supplementary Information). Single-cycle helical THz pulses are inherently elliptically polarized, but for simplicity we label them as left circularly polarized (LCP) and right circularly polarized (RCP). The ionic dynamics were detected by transient SHG intensity $I_{SHG}(\sigma^+, \sigma^-)$ with two opposite circular polarizations of the probing light ($\sigma^+$ and $\sigma^-$), which is sensitive to time-reversal symmetry breaking[38-40] due to the distinct second-order susceptibility tensor of $\sigma^+$ and $\sigma^-$. An analyzer whose polarization axis was aligned with the [010] crystal direction of STO was used to simplify the analysis of the SHG signal.

Figures 2a and 2b show $I_{SHG}(\sigma^+)$ (blue) and $I_{SHG}(\sigma^-)$ (red) obtained at 10 K in the QPE phase of STO under RCP and LCP THz excitation, respectively. The dominant effect of both the RCP and the LCP THz pulses is a non-oscillatory signal that exponentially decays with the relaxation time of 10 ps, which is attributed to transient inversion symmetry breaking[41]. We note that the intensity of the non-oscillating component for the elliptically polarized THz pump is about 80% of that for the vertically linearly polarized pump without any THz helicity-dependence (Section S2, Supplementary Information). The difference can be accounted for by different *E*-field strengths, indicating that the non-oscillatory component is unrelated to THz helicity.

In addition, we observe prominent fast oscillations at short times $t < 3$ ps, and slow oscillations at $t > 3$ ps, showing significant probe-helicity and THz-helicity dependencies. The fast oscillatory SHG signals are observed under both elliptically and linearly polarized THz excitation, and are similar to the previously reported THz-field-induced transient oscillations in the QPE phase[41]. In contrast, the slow oscillatory SHG signal disappears under linearly polarized THz excitation, suggesting that the circular polarization of the THz field is essential for initiation of the dynamics that produce this signal (Section S3, Supplementary Information)

SHG-CD highlights oscillatory signals that are odd under time reversal. Under THz excitation, transient SHG signals can be attributed to symmetry breaking by the THz-induced polarization[41] and axial phonons, and hot phonon effect[42]. The THz-induced polarization leads to symmetry lowering from 4/*mmm* to monoclinic *m*, and axial TO phonons break the time-reversal symmetry, further lowering the symmetry from *m* to *m′* magnetic point group. For *m′* symmetry, our analysis shows that only time non-invariant susceptibility tensors contribute to a finite SHG-CD signal $I_{CD} = I_{SHG}(\sigma^+) - I_{SHG}(\sigma^-)$ (Supplementary Note 1, Supplementary Information). Both the fast and the slow oscillatory signals in Figs. 2a and 2b depend on the probe polarization, resulting in a non-zero $I_{CD}$. Figure 2c shows $I_{CD}$ obtained under RCP (purple) and LCP (orange) THz excitation. The non-oscillatory signals attributed to inversion symmetry breaking are cancelled and only the time-reversal-symmetry-broken fast and the slow oscillatory signals are observed. This is confirmed by the absence of the dependence on the THz helicity for linear probe polarization that is sensitive to the inversion symmetry breaking (Section S2, Supplementary Information).

The fast oscillatory signals reverse sign with the THz helicity, consistent with their origin from time-reversal symmetry breaking. On the other hand, the slow signals do not fully reverse between RCP and LCP THz pumps, which can be attributed to imperfect circularity of the probe and/or a slight asymmetry between the profiles of opposite-helicity THz pulses (Section S4, Supplementary Information). The pure THz-helicity-dependent $\Delta I_{CD} = I_{CD}(RCP) - I_{CD}(LCP)$ between the signals for the two THz helicities can eliminate these artifact effects. As shown in the inset of Fig. 2c, $\Delta I_{CD}$ clearly shows two oscillating signals, a fast short-term transient and a slow oscillation at longer times, indicating the time-reversal symmetry breaking under elliptically polarized THz excitation. We used additional measurements of linear dichroism to independently confirm these behaviors and verify the origin of the oscillatory signals from THz helicity-dependence (Section S5, Supplementary Information).

Non-zero $\Delta I_{CD}$ signals for the fast oscillation can also arise from the third-order THz-field-induced SHG (TFISH) even in the absence of time-reversal symmetry breaking (Supplementary Note 2, Supplementary Information). In our experimental configuration, the TFISH signal is expected to be linear in the THz field $E_{THz}$ due to heterodyning with the background SHG[42]. However, the fast oscillatory signal exhibits a significantly different response in both the time and frequency domains compared to $E_{\mathrm{THz}}$ (Section S6, Supplementary Information). This suggests that the TFISH contribution in the QPE phase is negligible.

Significant temperature-dependence in both fast and slow oscillations indicates the quantum mechanical origin rather than classical effect. Figure 3a displays THz-helicity-dependent $\Delta I_{\mathrm{CD}}$ obtained at various temperatures. The peak positions from both the fast and slow oscillations were obtained by fitting the spectra with Lorentzian functions (Fig. 3c). The FFT spectra of the time-domain data show that the frequency $f_{\mathrm{H}} \approx 1$ THz of the fast oscillation is close to $2f_{\mathrm{TO}}$ and increases upon heating, as shown in Fig. 3b. Meanwhile, the frequency $f_{\mathrm{L}} \approx 0.2$ THz of the slow oscillation decreases with increasing temperature and cannot be directly attributed to a low-frequency mode such as the transverse acoustic (TA) phonon[43]. In particular, $f_{\mathrm{L}}$ exhibits transition-like behavior that follows the equation $\sim (1-T/T^{*})^{0.5}$ (red line in Fig. 3c), with $T^{*} \approx$ 60 K, somewhat higher than the QPE transition temperature ($T_{\mathrm{QPE}} \approx 40$ K). Instead, $T^{*}$ lies at the temperature associated with the formation[44-48] or evolution[49] of polar nanodomains that remain the subject of active theoretical modeling as well as experimental study[50,51].

In the classical picture, both the fast and slow ionic magnetic oscillation can be induced by dynamic multiferroicity ($\boldsymbol{M} \sim \boldsymbol{p} \times \mathrm{d}\boldsymbol{p}/\mathrm{d}t$)[52] from the beating between two polarized TO phonon modes under inversion symmetry breaking or the emergent new TO mode due to AFD tetragonal transition below about 100 K. In this case, the relaxation times of both polarization and magnetization oscillations should be identical. However, the slow magnetic oscillations observed in SHG-CD exhibit a much longer relaxation time (~8.9 ps) than the polarization oscillations (~1.7 ps), which cannot be explained by the classical mechanism. Furthermore, the slow oscillatory signal disappears at the crossover from the QPE to the classical PE state at 60 K, demonstrating the central role of the QPE state rather than the AFD state. Thus, our results represent the first direct observation, to our knowledge, of dynamic quantum ionic magnetism in the QPE state. A far weaker effect has recently been observed in room temperature STO, arising from a different mechanism through non-resonant THz pumping[53].

Quantum dynamics driven by linearly polarized THz pulses has been modelled in terms of the eigenstates of double-well ionic potential[34]. To describe two-dimensional ion dynamics, we consider the quantum ionic states in the four-well displacement potential reflecting the four-fold symmetry of STO in the (001) sample plane. We note that the ground state $\psi_0$ is symmetric, while the degenerate lowest-energy excited states $\psi_x$ and $\psi_y$ with energies $\hbar\omega_x$ and $\hbar\omega_y$ are antisymmetric with respect to the *x*- and *y*-axes, respectively (Fig. 4a and Supplementary Note 3, Supplementary Information).

Resonant elliptically polarized THz excitation breaks both inversion and time-reversal symmetry, lifting the degeneracy between the $\psi_x$ and $\psi_y$ states and leading to quantum beating between the resulting nondegenerate states (Fig. 4b). In the ferroelectric state of $BaTiO_3$, the $A_{1u}$ mode frequency is 4-6 times than that of the $E_u$ mode[54], so a large splitting comparable to the mode frequency is expected due to the THz-field-induced polarization[41]. The single-ion polarization $\boldsymbol{p}_i$ is determined by the coherence between the ground and the excited states $\boldsymbol{p}_i \propto |\psi\rangle\langle\psi_0| + h.c.$, where $\psi = (\psi_x, \psi_y)$, resulting in oscillation of the two components at the frequencies $\omega_x$ and $\omega_y$, respectively. On the other hand, the magnetic moment is determined by the amplitudes of the superposed excited state $\psi_\pm = (\psi_x \pm i\psi_y)/\sqrt{2}$ carrying circulating ionic currents, $M_z \propto |\psi_+\rangle\langle\psi_+| - |\psi_-\rangle\langle\psi_-| = 2i|\psi_y\rangle\langle\psi_x| + |\psi_x\rangle\langle\psi_y|$. Since the coherence governing magnetization is independent from that governing polarization, their relaxation rates are generally different. Our measurements demonstrate that magnetization oscillation decay significantly more slowly than polarization oscillations. This is consistent with viscous-like relaxation mechanisms, due to the larger frequency of the latter. While the polarization oscillation rapidly dephases due to the large energy separation between the ground and excited states $\hbar\omega_{x,y}$, the magnetization relaxes on a much slower timescale due to the small energy splitting $\hbar(\omega_x - \omega_y)$. As a result, the quantum ionic magnetization decays more slowly than the polarization, which is distinct from a classical effect.

To confirm this analysis, we performed numerical simulations based on the Lindblad master equation for the density matrix in the basis of states $\psi_0$, $\psi_x$, and $\psi_y$. We included the symmetry-breaking effect of non-oscillatory transient polarization $\boldsymbol{p}_n$ whose dynamics was simulated by the Landau-Khalatnikov equation. The polarization-dependent TO mode stiffening was modelled by the dependence $\omega_i(p_{n,i}) \approx (1 + \alpha p_{n,i}^2)\omega_0$, where $i = x, y$, reflecting the inversion symmetry of STO (Supplementary Note 3, Supplementary Information). This results in the oscillation of $M_z$ at the frequency $\Delta\omega = \alpha|p_{n,x}^2 - p_{n,y}^2|\omega_0$ determined by the THz-induced symmetry breaking between quantum ionic states.

As shown in Fig. 4c, the simulation reproduces the non-oscillatory gradually decaying component $p_{n,i}$ (top) and the rapidly relaxing oscillatory polarization $p_{osc,i}$ (middle), consistent with $I_{SHG}$ measured under linear probe polarization (Section S2, Supplementary Information). The magnetization $M_z$ (bottom) shows fast initial oscillations due to the rapid modulations of the components $\psi_x$ and $\psi_y$ driven by the THz pulse, followed by slow oscillations. The time-dependent $M_z$ reverses sign on reversing the helicity of the THz pulse, consistent with our observations. Notably, the slow oscillations persist after the polarization oscillations disappear, indicating the quantum mechanical origin (dashed lines in the middle and bottom panels). This result is in good agreement with the experimental observations in Fig. 2c. We note that the frequency of the slow oscillation is softened as the amplitude of THz pulse decreases, due to the reduction of $\hbar(\omega_x - \omega_y)$ (Section S7, Supplementary Information). This models the effects of frequency softening of slow oscillations observed in Fig. 3b due to the reduction of inversion symmetry breaking upon heating (Section S8, Supplementary Information). Our consistent experimental and theoretical results demonstrate ultrafast control of ionic magnetization by THz-induced inversion symmetry breaking. Although non-*c*-axis AFD domains can in principle introduce nondegenerate soft modes, the observed beating is unlikely to originate from static domain anisotropy because it vanishes at much lower than AFD transition temperature (~ 105 K), at which the energy difference of nondegenerate soft modes continuously softens to zero[55].

In summary, we showed that elliptically polarized, resonant THz pulses induce ultrafast ionic magnetic oscillations in QPE $SrTiO_3$. These oscillations persist after the polarization oscillations disappear and are observed only in the low-temperature where the quantum fluctuation is dominant, demonstrating their quantum origin. Analysis confirms that the oscillations originate from the quantum beating between dynamical states whose degeneracy is lifted by the THz-induced polarization. Our findings provide a novel pathway to control ultrafast magnetic oscillations in non-magnetic quantum materials, and offer new opportunities for THz spintronic and magnetic quantum device applications.

## Methods

### Generation of elliptically polarized THz pulse

We generated elliptically polarized THz pulses by introducing a time delay between two orthogonal THz *E*-field components[53]. A vertically polarized THz pulse was generated using a $LiNbO_3$ prism via the tilted-pulse-front method with a 1 kHz amplified pulsed laser (Astrella, *Coherent*) with 800 nm center wavelength and 5 mJ pulse energy. The peak amplitude of the THz field was about 300 KV/cm, as estimated from the THz power and beam size, and the center frequency was 0.5 THz. A wire-grid polarizer was used to separate the vertically polarized THz pulse into two components polarized at +45 and –45 degrees. By varying the time delay between the two components, elliptically polarized THz pulses with opposite helicities were produced.

### THz-pump-induced second harmonic generation

To obtain second harmonic generation (SHG) signals to investigate the ultrafast quantum ionic magnetization in STO, the fundamental light with center wavelength of 800 nm was passed through a 600 nm long-pass optical filter to block any second-harmonic light from the optics. The polarization was set along the vertical direction using a Glan-Taylor polarizer with extinction ratio > 3000. A zeroth-order quarter-wave-plate (QWP) with the center wavelength of 800 nm was used to produce circularly polarized fundamental light, and the left-handed and right-handed polarization states were carefully calibrated using a polarizer-QWP-analyzer polarimetry method. We detected second-harmonic light from STO at normal incidence in a transmission configuration using a photomultiplier tube, with the fundamental light blocked by a 400 nm band-pass optical filter. By modulating the THz pulse using a 500 Hz mechanical chopper, THz-pump-induced SHG signals were detected using a lock-in amplifier (SR830, *SRS*). Cryogenic measurements were performed using a flow-type cryostat (*Lakeshore*) with TPX windows.

## Acknowledgement

This work was supported in part by the Office of Naval Research (grant no. 14035670). I.H.C. was supported by the National Research Foundation of Korea (NRF) grant funded by the Korea government (MSIT) (No. RS-2024-00351794). S.U. was supported by the US National Science Foundation (NSF) award No. ECCS- 2448290

## Contributions

I.H.C., S.U., and K.A.N. conceived the idea and designed the experiments. I.H.C performed temperature-dependent SHG measurement and symmetry analysis. I.H.C. performed temperature-dependent LD experiments. S.U. performed numerical simulations. I.H.C., S.U. and K.A.N. drafted the manuscript initially and revised it based on input and feedback from all the authors. K.A.N. directed this project.

## Corresponding author

Correspondence to Keith A. Nelson.

## Competing interests

The authors declare no competing interests.

## Data availability

All data supporting the results within this paper and supporting information are available from the corresponding author upon reasonable request. Source data are provided with this paper.

## Code availability

All numerical simulation codes employed this work are available from the corresponding author upon reasonable request.

## Figures

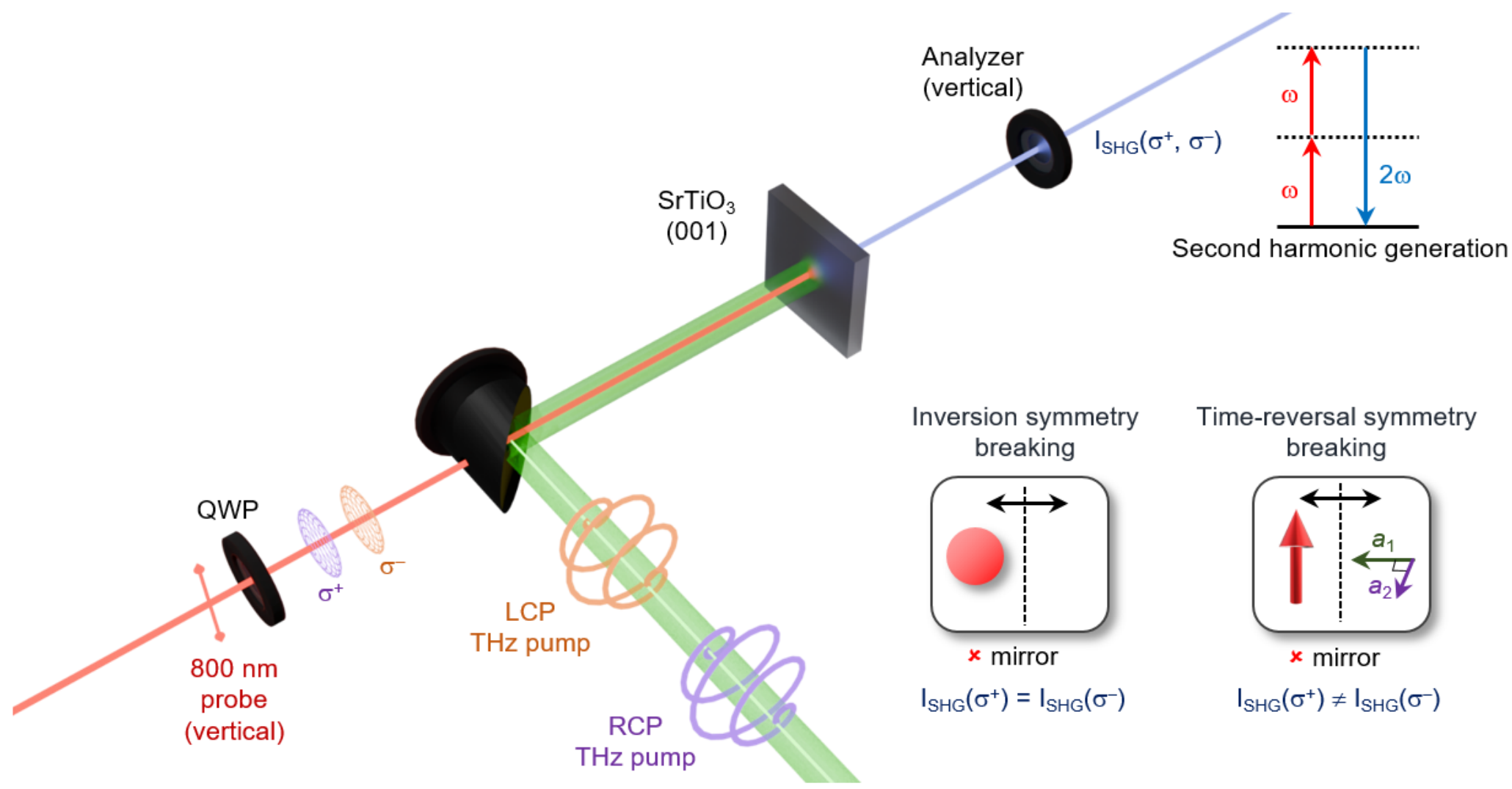

**Figure 1. THz-field-driven ultrafast manipulation of polarization and magnetization in quantum paraelectric $SrTiO_3$.** Elliptically polarized THz excitation can break both inversion and time-reversal symmetry, leading to transient polarization and magnetization in $SrTiO_3$. To monitor the time-reversal symmetry breaking, second harmonic generation circular dichroism (SHG-CD) was measured using circularly polarized optical pulses with two helicities ($\sigma^+$ and $\sigma^-$) under left-handed (LCP) and right-handed (RCP) THz excitation.

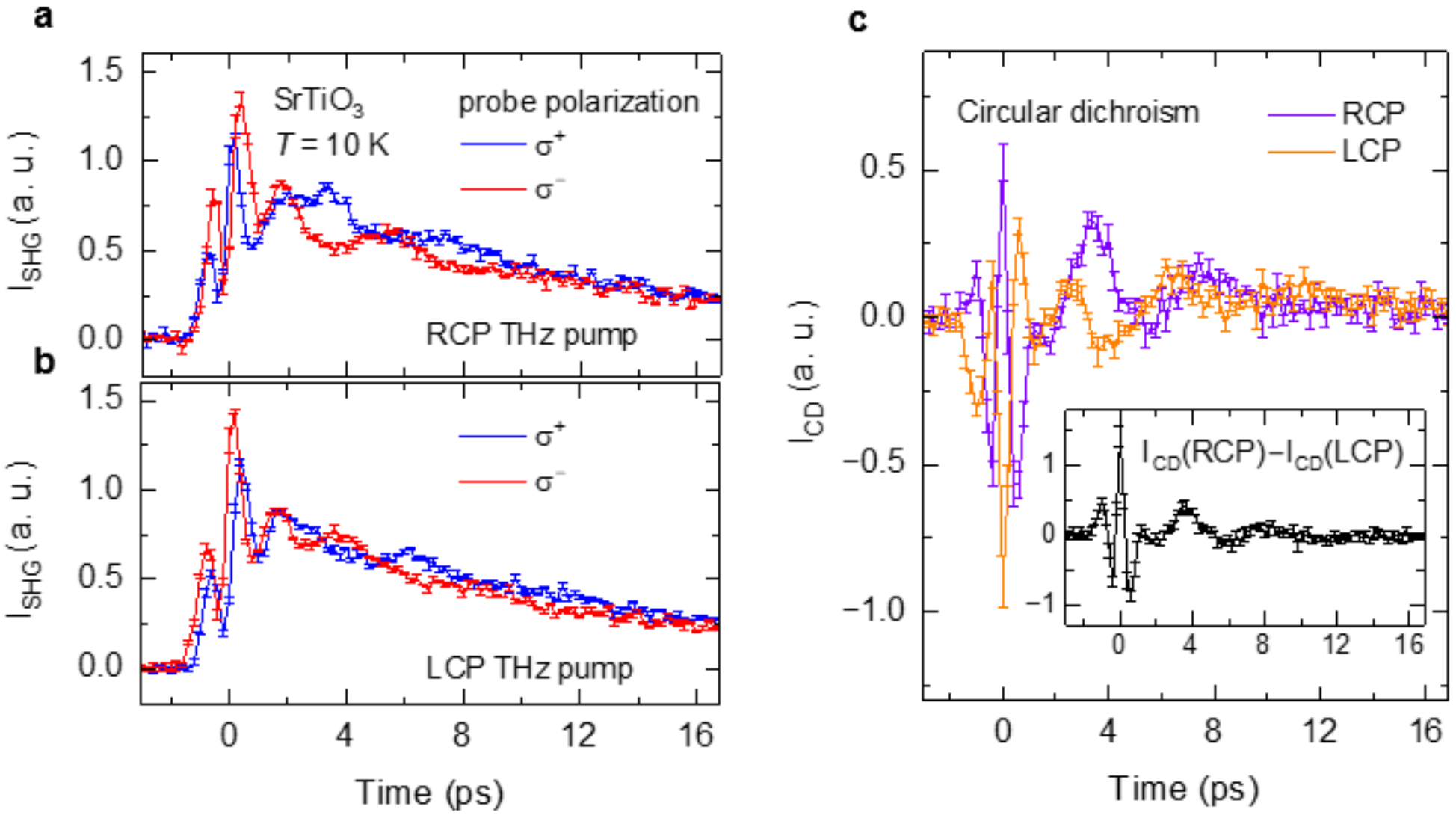

**Figure 2. Time-reversal-symmetry-broken THz oscillations in second harmonic generation circular dichroism. (a, b)** Transient SHG signals $I_{SHG}$ at 10 K obtained with right-handed ($\sigma^+$, blue) and left-handed ($\sigma^-$, red) circular probe polarizations under right-handed (RCP) and left-handed (LCP) THz excitation. **(c)** Transient SHG-CD signals $I_{CD}$ under RCP (purple) and LCP (orange) THz excitation. The inset shows the $I_{CD}$ difference between the two THz helicities.

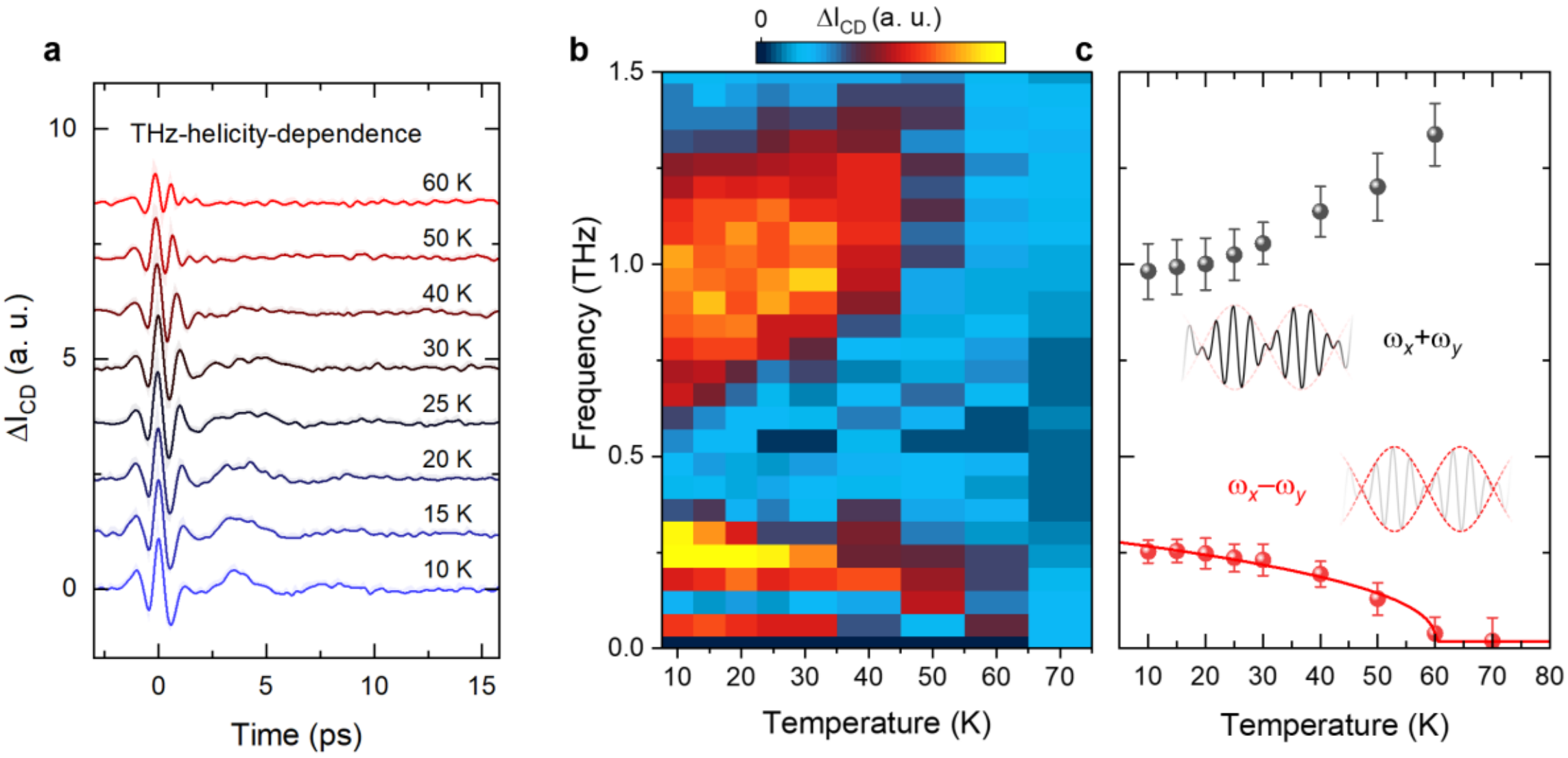

**Figure 3. Temperature-dependent fast and slow oscillatory signals in elliptically polarized THz-helicity-dependent second harmonic generation circular dichroism. (a)** Transient $\Delta I_{CD}$ obtained at various temperatures. **(b)** Temperature-dependent FFT spectrum of $\Delta I_{CD}$, showing two distinct peaks at around 0.2 and 1 THz with significant temperature dependence. **(c)** Temperature-dependent peak positions, attributed to ionic magnetization induced by quantum beating between the *x*- and *y*-direction axial transverse optical (TO) phonon oscillations at frequencies $\omega_x$ and $\omega_y$, respectively. The red line represents a fit to the transition-like behavior, $\sim (1-T/T^*)^{0.5}$, where $T^* \approx 60$ K.

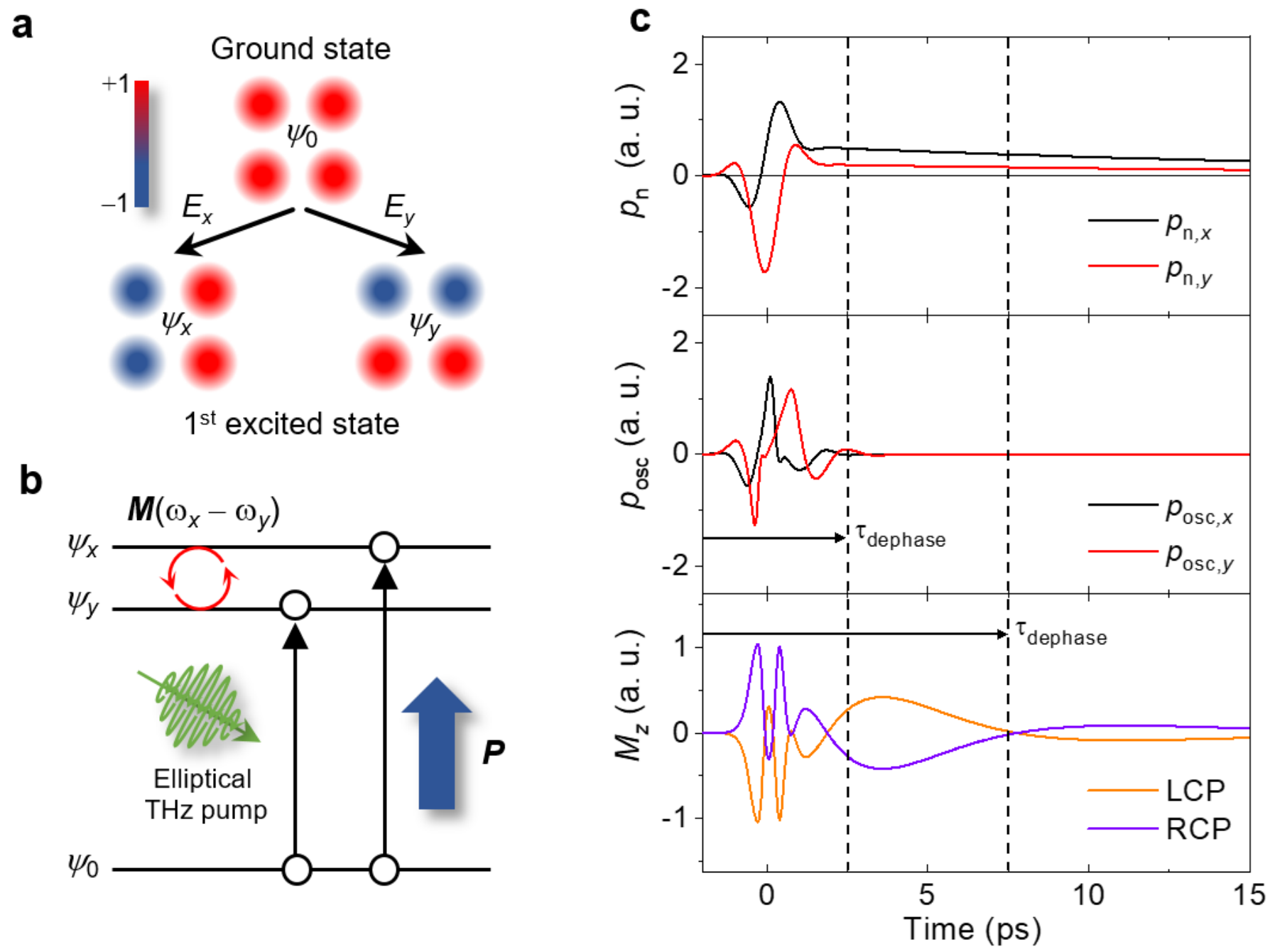

**Figure 4. Numerical simulation for elliptically polarized THz-pump-induced quantum ionic polarization and magnetization. (a)** Ground and first-excited quantum ionic states in a four-well displacement potential in the STO (001) plane. **(b)** Elliptically polarized THz pump-induced inversion and time-reversal symmetry breaking, resulting in macroscopic polarization $\boldsymbol{P}$ and magnetization $\boldsymbol{M}$. **(c)** Transient in-plane ($i = x, y$) non-oscillatory polarization $p_{n,i}$ (top), oscillatory polarization $p_{osc,i}$ (middle), and the out-of-plane magnetization $M_z$ (bottom). Dashed lines indicate $\tau_{dephase}$ of polarization and slow magnetization oscillations.

## References

1 Zhu, H., Yi, J., Li, M.-Y., Xiao, J., Zhang, L., Yang, C.-W., Kaindl, R. A., Li, L.-J., Wang, Y., and Zhang, X. Observation of chiral phonons. *Science* **359**, 579 (2018).

2 Yin, T., Ulman, K. A., Liu, S., Granados del Águila, A., Huang, Y., Zhang, L., Serra, M., Sedmidubsky, D., Sofer, Z., and Quek, S. Y. Chiral phonons and giant magneto-optical effect in $CrBr_3$ 2D magnet. *Advanced Materials* **33**, 2101618 (2021).

3 Cui, J., Boström, E. V., Ozerov, M., Wu, F., Jiang, Q., Chu, J.-H., Li, C., Liu, F., Xu, X., and Rubio, A. Chirality selective magnon-phonon hybridization and magnon-induced chiral phonons in a layered zigzag antiferromagnet. *Nature Communications* **14**, 3396 (2023).

4 Ishito, K., Mao, H., Kobayashi, K., Kousaka, Y., Togawa, Y., Kusunose, H., Kishine, J. i., and Satoh, T. Chiral phonons: circularly polarized Raman spectroscopy and ab initio calculations in a chiral crystal tellurium. *Chirality* **35**, 338 (2023).

5 Ishito, K., Mao, H., Kousaka, Y., Togawa, Y., Iwasaki, S., Zhang, T., Murakami, S., Kishine, J.-i., and Satoh, T. Truly chiral phonons in α-HgS. *Nature Physics* **19**, 35 (2023).

6 Ueda, H., García-Fernández, M., Agrestini, S., Romao, C. P., van den Brink, J., Spaldin, N. A., Zhou, K.-J., and Staub, U. Chiral phonons in quartz probed by X-rays. *Nature* **618**, 946 (2023).

7 Oishi, E., Fujii, Y., and Koreeda, A. Selective observation of enantiomeric chiral phonons in α-quartz. *Physical Review B* **109**, 104306 (2024).

8 Kim, C., Hwang, I. K., Moon, K. W., An, K., Lee, K. J., Ko, J. H., Park, B. G., Choi, K. Y., and Hwang, C. Chiral Acoustic Phonon and Conservation of Pseudoangular Momentum in α-Quartz. *Advanced Materials*, e11289 (2025).

9 Wu, F., Zhou, J., Bao, S., Li, L., Wen, J., Wan, Y., and Zhang, Q. Magnetic switching of phonon angular momentum in a ferrimagnetic insulator. *Physical Review Letters* **134**, 236701 (2025).

10 Vitale, S. A., Nezich, D., Varghese, J. O., Kim, P., Gedik, N., Jarillo-Herrero, P., Xiao, D., and Rothschild, M. Valleytronics: opportunities, challenges, and paths forward. *Small* **14**, 1801483 (2018).

11 Yang, S.-H., Naaman, R., Paltiel, Y., and Parkin, S. S. Chiral spintronics. *Nature Reviews Physics* **3**, 328 (2021).

12 Davies, C., Fennema, F., Tsukamoto, A., Razdolski, I., Kimel, A., and Kirilyuk, A. Phononic switching of magnetization by the ultrafast Barnett effect. *Nature* **628**, 540 (2024).

13 Luo, J., Lin, T., Zhang, J., Chen, X., Blackert, E. R., Xu, R., Yakobson, B. I., and Zhu, H. Large effective magnetic fields from chiral phonons in rare-earth halides. *Science* **382**, 698 (2023).

14 Basini, M., Pancaldi, M., Wehinger, B., Udina, M., Unikandanunni, V., Tadano, T., Hoffmann, M. C., Balatsky, A. V., and Bonetti, S. Terahertz electric-field-driven dynamical multiferroicity in $SrTiO_3$. *Nature* **628**, 534 (2024).

15 Nova, T. F., Cartella, A., Cantaluppi, A., Först, M., Bossini, D., Mikhaylovskiy, R. V., Kimel, A. V., Merlin, R., and Cavalleri, A. An effective magnetic field from optically driven phonons. *Nature Physics* **13**, 132 (2017).

16 Zhang, H., Peshcherenko, N., Yang, F., Ward, T. Z., Raghuvanshi, P., Lindsay, L., Felser, C., Zhang, Y., Yan, J.-Q., and Miao, H. Measurement of phonon angular momentum. *Nature Physics* **21**, 1387 (2025).

17 Wefers, M. M., Kawashima, H., and Nelson, K. A. Optical control over two-

dimensional lattice vibrational trajectories in crystalline quartz. *The Journal of Chemical Physics* **108**, 10248 (1998).
18 Fleischer, S., Zhou, Y., Field, R. W., and Nelson, K. A. Molecular orientation and alignment by intense single-cycle THz pulses. *Physical Review Letters* **107**, 163603 (2011).
19 Tauchert, S. R., Volkov, M., Ehberger, D., Kazenwadel, D., Evers, M., Lange, H., Donges, A., Book, A., Kreuzpaintner, W., and Nowak, U. Polarized phonons carry angular momentum in ultrafast demagnetization. *Nature* **602**, 73 (2022).
20 Dornes, C., Acremann, Y., Savoini, M., Kubli, M., Neugebauer, M. J., Abreu, E., Huber, L., Lantz, G., Vaz, C. A., and Lemke, H. The ultrafast Einstein–de Haas effect. *Nature* **565**, 209 (2019).
21 Choi, I. H., Jeong, S. G., Song, S., Park, S., Shin, D. B., Choi, W. S., and Lee, J. S. Real-time dynamics of angular momentum transfer from spin to acoustic chiral phonon in oxide heterostructures. *Nature nanotechnology* **19**, 1277 (2024).
22 Cheng, B., Schumann, T., Wang, Y., Zhang, X., Barbalas, D., Stemmer, S., and Armitage, N. A large effective phonon magnetic moment in a Dirac semimetal. *Nano letters* **20**, 5991 (2020).
23 Baydin, A., Hernandez, F. G., Rodriguez-Vega, M., Okazaki, A. K., Tay, F., Noe, G. T., Katayama, I., Takeda, J., Nojiri, H., and Rappl, P. H. Magnetic control of soft chiral phonons in PbTe. *Physical Review Letters* **128**, 075901 (2022).
24 Juraschek, D. M., Narang, P., and Spaldin, N. A. Phono-magnetic analogs to opto-magnetic effects. *Physical Review Research* **2**, 043035 (2020).
25 Juraschek, D. M., Neuman, T., and Narang, P. Giant effective magnetic fields from optically driven chiral phonons in 4*f* paramagnets. *Physical Review Research* **4**, 013129 (2022).
26 Geilhufe, R. M., Juričić, V., Bonetti, S., Zhu, J.-X., and Balatsky, A. V. Dynamically induced magnetism in $KTaO_3$. *Physical Review Research* **3**, L022011 (2021).
27 Urazhdin, S. Atomic and interatomic orbital magnetization induced in $SrTiO_3$ by chiral phonons. *Physical Review B* **111**, 214435 (2025).
28 Kim, K., Vetter, E., Yan, L., Yang, C., Wang, Z., Sun, R., Yang, Y., Comstock, A. H., Li, X., and Zhou, J. Chiral-phonon-activated spin Seebeck effect. *Nature Materials* **22**, 322 (2023).
29 Ohe, K., Shishido, H., Kato, M., Utsumi, S., Matsuura, H., and Togawa, Y. Chirality-induced selectivity of phonon angular momenta in chiral quartz crystals. *Physical Review Letters* **132**, 056302 (2024).
30 Chen, H., Wu, W., Zhu, J., Yang, Z., Gong, W., Gao, W., Yang, S. A., and Zhang, L. Chiral phonon diode effect in chiral crystals. *Nano letters* **22**, 1688 (2022).
31 Choi, I. H., Kim, D., Jin, Y. J., Yang, S., Ju, T.-S., Kim, C., Hwang, C., Shin, D., and Lee, J. S. Demagnetization-Driven Nanoscale Chirality-Selective Thermal Switch. *arXiv:2509.24205*(2025).
32 Shin, D., Latini, S., Schäfer, C., Sato, S. A., De Giovannini, U., Hübener, H., and Rubio, A. Quantum paraelectric phase of $SrTiO_3$ from first principles. *Physical Review B* **104**, L060103 (2021).
33 Monkhorst, H. J., and Pack, J. D. Special points for Brillouin-zone integrations. *Physical Review B* **13**, 5188 (1976).
34 Shin, D., Latini, S., Schäfer, C., Sato, S. A., Baldini, E., De Giovannini, U., Hübener, H., and Rubio, A. Simulating terahertz field-induced ferroelectricity in quantum paraelectric $SrTiO_3$. *Physical Review Letters* **129**, 167401 (2022).
35 Müller, K. A., Berlinger, W., and Tosatti, E. Indication for a novel phase in the quantum

paraelectric regime of $SrTiO_3$. *Zeitschrift für Physik B Condensed Matter* **84**, 277 (1991).
36 Aschauer, U., and Spaldin, N. A. Competition and cooperation between antiferrodistortive and ferroelectric instabilities in the model perovskite $SrTiO_3$. *Journal of Physics: Condensed Matter* **26**, 122203 (2014).
37 Marqués, M. I., Aragó, C., and Gonzalo, J. A. Quantum paraelectric behavior of $SrTiO_3$: Relevance of the structural phase transition temperature. *Physical Review B—Condensed Matter and Materials Physics* **72**, 092103 (2005).
38 Petralli-Mallow, T., Wong, T., Byers, J., Yee, H., and Hicks, J. Circular dichroism spectroscopy at interfaces: a surface second harmonic generation study. *The Journal of Physical Chemistry* **97**, 1383 (1993).
39 Byers, J., Yee, H., Petralli-Mallow, T., and Hicks, J. Second-harmonic generation circular-dichroism spectroscopy from chiral monolayers. *Physical Review B* **49**, 14643 (1994).
40 Fiebig, M., Pavlov, V. V., and Pisarev, R. V. Second-harmonic generation as a tool for studying electronic and magnetic structures of crystals. *Journal of the Optical Society of America B* **22**, 96 (2005).
41 Li, X., Qiu, T., Zhang, J., Baldini, E., Lu, J., Rappe, A. M., and Nelson, K. A. Terahertz field–induced ferroelectricity in quantum paraelectric $SrTiO_3$. *Science* **364**, 1079 (2019).
42 Yang, F., Li, X., Talbayev, D., and Chen, L. Terahertz-induced second-harmonic generation in quantum paraelectrics: hot-phonon effect. *Physical Review Letters* **135**, 056901 (2025).
43 Orenstein, G., Krapivin, V., Huang, Y., Zhang, Z., de la Peña Muñoz, G., Duncan, R. A., Nguyen, Q., Stanton, J., Teitelbaum, S., and Yavas, H. Observation of polarization density waves in $SrTiO_3$. *Nature Physics* **21**, 961 (2025).
44 Burns, G., and Dacol, F. Crystalline ferroelectrics with glassy polarization behavior. *Physical Review B* **28**, 2527 (1983).
45 Burns, G., and Dacol, F. Soft phonons in a ferroelectric polarization glass system. *Solid state communications* **58**, 567 (1986).
46 Burns, G., and Dacol, F. Glassy polarization behavior in $K_2Sr_4(NbO_3)_{10}$-type ferroelectrics. *Physical Review B* **30**, 4012 (1984).
47 Burns, G., Dacol, F., and Taylor, W. Optical properties of ferroelectric $(Pb_{1-x}Ba_x)_5Ge_3O_{11}$ for $x = 0$ and 0.02. *Physical Review B* **28**, 2531 (1983).
48 Wang, C., and Zhao, M. Burns temperature and quantum temperature scale. *Journal of Advanced Dielectrics* **1**, 163 (2011).
49 Zhang, Y., Sung, S. H., Agarwal, N., Gates, M., Li, C., Yu, P., Hovden, R., and El Baggari, I. Imaging of nanoscale polar textures in quantum paraelectric $SrTiO_3$. *Nature* **656**, 54 (2026).
50 Bussmann-Holder, A., Kremer, R. K., Roleder, K., and Salje, E. K. H. $SrTiO_3$: Thoroughly Investigated but Still Good for Surprises. *Condensed Matter* **9**, 3 (2024).
51 Sun, Y.-J., Yang, F., and Chen, L.-Q. Polar nanoregions and reentrant-like ferroelectric behavior in $SrTiO_3$. *arXiv:2609.01446*(2026).
52 Juraschek, D. M., Fechner, M., Balatsky, A. V., and Spaldin, N. A. Dynamical multiferroicity. *Physical Review Materials* **1**, 014401 (2017).
53 Choi, I. H., Urazhdin, S., Varshney, S., Jeong, S. G., Jalan, B., and Nelson, K. A. Light-Driven Ultrafast Control of Time-Reversal Symmetry in $SrTiO_3$. *arXiv:2609.09497* (2026).
54 Wan, F., Han, J., and Zhu, Z. Dielectric response in ferroelectric $BaTiO_3$. *Physics Letters A* **372**, 2137 (2008).

55 Yamanaka, A., Kataoka, M., Inaba, Y., Inoue, K., Hehlen, B., and Courtens, E. Evidence for competing orderings in strontium titanate from hyper-Raman scattering spectroscopy. *EPL (Europhysics Letters)* **50**, 688 (2000).

# Supporting Information

## THz-Driven Quantum Ionic Magnetism in a Quantum Paraelectric $SrTiO_3$

In Hyeok Choi[1,†], Sergei Urazhdin[2], Man Tou Wong[1], Zi-Jie Liu[1], and Keith A. Nelson[1,*]

[1]*Department of Chemistry, Massachusetts Institute of Technology, Cambridge, Massachusetts 02139, United states*

[2]*Department of Physics, Emory University, Atlanta, Georgia 30322, United states*

[†] First author

[*]Corresponding authors: kanelson@mit.edu

**Supplementary Note 1 – Analytic solution for SHG circular dichroism**

In its antiferrodistortive phase ($T$ < 105 K), $SrTiO_3$ (STO) exhibits the centrosymmetric I4/*mcm* space group[1], which contains four-fold and four-fold-screw rotations, a two-fold rotation, four glide planes, and mirror planes oriented both parallel and perpendicular to the *z* axis. Under elliptically polarized THz excitation, THz-field-induced polarization emerge along both the *x*- and *y*-directions with different amplitudes due to the ellipticity of THz pump. In this case, polarization breaks the mirror symmetry planes perpendicular to the polarization, leading to symmetry lowering into non-centrosymmetric monoclinic *m*. If time-reversal-symmetry-broken axial phonons are induced by elliptically polarized THz pumping, the symmetry of STO is further lowered into $m'$.

Considering non-centrosymmetric magnetic point group $m'$, second harmonic light can be induced by the electric dipole (ED) process, with $E_j(2\omega) = \chi^{(2)}_{jkl} E_k(\omega) E_l(\omega)$, where $\chi^{(2)}_{jkl}$ are second-order susceptibility tensor components. For a circularly polarized optical pulse, $E(\omega)$ can have both real and imaginary values, $E_x \pm iE_y$. Under broken time-reversal symmetry, $\chi^{(2)}_{jkl}$ can be consist of time invariant (*i*-tensor, $\chi^{i}_{jkl}$) and time non-invariant (*c*-tensor, $\chi^{c}_{jkl}$) tensor components, which are even and odd under time-reversal. We obtained the non-zero *i*- and *c*-tensor components satisfying the relation[2],

$$\chi^{i}_{jkl} = \sigma_{jp}\sigma_{kq}\sigma_{lr}\chi^{i}_{pqr}, \qquad \chi^{c}_{jkl} = |S|\sigma_{jp}\sigma_{kq}\sigma_{lr}\chi^{c}_{pqr}, \tag{S1}$$

where σ is a crystalline symmetry operator, and $S$ is a time-reversal symmetry operator. We note that *i*-tensor and *c*-tensor are pure real and imaginary, respectively due to the time-reversal symmetry argument[3]. We summarize in Table S1 the non-zero components for both *i*- and *c*-tensors and the corresponding third-order susceptibility components that yield a *y*-polarized second-harmonic *E*-field with right-handed (RCP), left-handed (LCP) and linear (LP) probe polarization. We note that the [010] axis is perpendicular to the output analyzer, and the light incidence angle is normal to the plane. Notably, SHG signals from *c*-tensors can only be observed for circular probe polarizations, while there are only structure-related *i*-tensor contributions for linear probe polarizations. Under alternating helicity of probe light, the second-harmonic *E*-field from *c*-tensors changes sign, resulting in a time-reversal-symmetry-broken non-zero SHG circular dichroism.

**Table S1. Non-zero tensor components for *i*- and *c*-tensors, and corresponding *y*-polarized second harmonic field with RCP, LCP, and LP probe polarization.**

| | | Time invariant (*i*-tensor) | Time non-invariant (*c*-tensor) |
|---|---|---|---|
| Non-zero components | | *yyy, yxx, yzz, yzy ,yyz, xxz, xzx, xyx, xxy, zyy ,zxx, zzz, zzy, zyz* | *yyx, yxy, yxz, yzx, xyy, xyz, xxx, xzy, xzz, zyx, zxy, zxz, zzx* |
| Probe polarization | RCP | $\chi^{i}_{yxx} - \chi^{i}_{yyy}$ | $2\chi^{c}_{yxy}$ |
| | LCP | $\chi^{i}_{yxx} - \chi^{i}_{yyy}$ | $-2\chi^{c}_{yxy}$ |
| | LP | $4\chi^{i}_{yyy}$ | 0 |

**Supplementary Note 2 – Analytic solution for THz-field-induced SHG**

Even in centrosymmetric materials, a THz electric field can enable optical second-harmonic generation through a third-order process known as THz-field-induced second harmonic generation (TFISH). The second harmonic field generated by TFISH can be written as $E_j(2\omega + \Omega) = \chi^{(3)}_{jklm} E_j^{THz}(\Omega) E_k(\omega) E_l(\omega)$, where $E^{\mathrm{THz}}$ is the electric field of a THz pulse, and $\Omega$ is its frequency that is much smaller than $\omega$. In Table S2, we summarize the analytic solutions of second harmonic field generation by TFISH for cubic $m3m$ and tetragonal $4/mmm$ point groups of room temperature and low temperature (< 105 K) STO, respectively, at normal incidence with RCP and LCP probe polarizations. Here we introduced a background SHG signal $E_0$, which can emerge even in centrosymmetric STO due to surface ED and bulk EQ processes[4,5]. Therefore, TFISH leads to non-zero circular dichroism (CD) signals $I_{\mathrm{CD}}$, which are given by $8\chi_{xxxx} E_0 E_x^{\mathrm{THz}}$ and $4(\chi_{yyxx} + \chi_{yxxy}) E_0 E_x^{\mathrm{THz}}$ for the cubic and tetragonal phase, respectively. When the THz helicity is reversed, the sign of $E_x^{\mathrm{THz}}$ flips while $E_y^{\mathrm{THz}}$ remains unchanged, resulting in a non-zero THz helicity-dependent contribution to the fast oscillatory signals. It is noteworthy that the THz-helicity-dependent TFISH-CD signals are simply proportional to $E_x^{\mathrm{THz}}$, and hence their FFT spectra should follow that of $E_x^{\mathrm{THz}}$. At room temperature, we confirmed that the FFT spectrum closely matches that of $E_x^{\mathrm{THz}}$, consistent with a TFISH origin. At low temperature, however, the spectrum shifts to a much higher center frequency and its peak amplitude increases by a factor of 30, indicating that an additional contribution, such as an ionic response, must be taken into account.

**Table S2. THz-field-induced second harmonic field with RCP, and LCP probe polarization for cubic *m*3*m* and tetragonal 4/*mmm* point group.**

| Symmetry | | Real | Imaginary |
|---|---|---|---|
| Cubic $m3m$ | RCP | $E_0 + 2\chi_{xxxx} E_y^{\mathrm{THz}}$ | $E_0 - 2\chi_{xxxx} E_x^{\mathrm{THz}}$ |
| | LCP | $E_0 + 2\chi_{xxxx} E_y^{\mathrm{THz}}$ | $E_0 + 2\chi_{xxxx} E_x^{\mathrm{THz}}$ |
| Tetragonal $4/mmm$ | RCP | $E_0 + (\chi_{yyyy} - \chi_{yxyx}) E_y^{\mathrm{THz}}$ | $E_0 - (\chi_{yyxx} + \chi_{yxxy}) E_x^{\mathrm{THz}}$ |
| | LCP | $E_0 + (\chi_{yyyy} - \chi_{yxyx}) E_y^{\mathrm{THz}}$ | $E_0 + (\chi_{yyxx} + \chi_{yxxy}) E_x^{\mathrm{THz}}$ |

**Supplementary Note 3. Simulations of oscillatory magnetization in QPE driven by an elliptically polarized THz pulse**

*Classical vs quantum THz-driven ionic dynamics in the QPE*

Our central experimental finding is the observation of an oscillatory magnetic SHG signal with frequency substantially below that of the TO mode, in addition to oscillation at about twice the TO mode frequency. The slow magnetic oscillation decreases in amplitude and redshifts as the temperature increases toward the quantum paraelectric (QPE) transition temperature $T_c$, while the high-frequency oscillation blueshifts.

In the classical picture of dynamical ionic multiferroicity, ionic magnetization results from the dynamical polarization[6], $\boldsymbol{M} \propto \boldsymbol{P} \times \dot{\boldsymbol{P}}$. However, we found that oscillating magnetic signal decays substantially more slowly than the contribution from the oscillatory polarization, which cannot be explained by the classical picture. Below, we first introduce a classical model of ionic polarization dynamics and magnetism, which captures some of our observations while confirming this limitation, and we subsequently develop a minimal quantum model elucidating the quantum mechanism of long-lived magnetism in the absence of polarization oscillation. This analysis shows that non-classical ionic states play an important role in both the static and dynamical properties of the QPE state in STO.

*Classical model of ionic polarization and magnetization*

The two essential features of transient SHG produced by the *linearly* polarized THz pulses in the QPE state of STO are i) non-oscillatory signal that decays on the time scale of about 10 ps at $T$ = 10 K, and ii) rapidly decaying oscillations of SHG at about 1 THz, approximately double the frequency $f_{TO} \approx 0.5$ THz of the soft TO phonon mode, which blue-shift with increasing $T$.

These features are reminiscent of the coexisting soft mode and non-oscillatory central mode in relaxor ferroelectrics (r-FE) described by two distinct order parameters[7]. Based on this similarity, we describe THz-induced ferroelectric ($t$-FE) polarization and the oscillating TO mode by two separate contributions to the dipole moment of the pseudo-cubic unit cell, $\boldsymbol{p}_{FE}$ and $\boldsymbol{p}_{TO}$, correspondingly. We emphasize that in contrast to relaxor ferroelectrics where $\boldsymbol{p}_{FE}$ originates from the rearrangements of FE-ordered domains, in QPE it originates from THz-induced net polarization resulting from transient metastable FE ordering. Following the established models of relaxor FEs, we model the dynamics of $\boldsymbol{p}_{\mathrm{FE}}$ using the Landau-Khalatnikov equation[8],

$$\tau_{\mathrm{FE}}\dot{\boldsymbol{p}}_{\mathrm{FE}} + \boldsymbol{p}_{\mathrm{FE}} = \frac{\chi_{\mathrm{FE}}}{v}\boldsymbol{E}(t), \tag{S2}$$

where $\tau_{\mathrm{FE}}$ is the relaxation time, $\chi_{\mathrm{FE}}$ is the contribution of the *t*-FE response to dc susceptibility, and $v$ is the volume of the unit cell. In this approximation, the field-free evolution of the *t*-FE polarization is exponential relaxation $\boldsymbol{p}_{\mathrm{FE}}(t) \propto e^{-t/\tau_{\mathrm{FE}}}$, in agreement with the time dependence of the non-oscillatory contribution to SHG observed in our measurements.

The dynamical TO dipolar moment per pseudo-cubic unit cell is $\boldsymbol{p}_{\mathrm{TO}} = Z^*Q$, where $Z^*$ is the effective mode charge and $Q$ is this mode's coordinate. Here and below, we normalize all the intensive quantities to the pseudo-cubic cell volume. In the Slater approximation for the TO mode, the Ti cation is displaced oppositely to the rigid oxygen octahedra, and Sr displacement is neglected. The mode coordinate is the difference between Ti and oxygen displacements, $\boldsymbol{Q} = \boldsymbol{u}_{\mathrm{Ti}} - \boldsymbol{u}_{\mathrm{O}}$, $Z^* = \sum_k \frac{Z_k^*}{\sqrt{\mathcal{M}_k}}\sqrt{\mathcal{M}} \approx 1.8$ is the mode charge, where the summation over the ions excludes Sr and the effective Born ion charges are $Z_{\mathrm{Ti}}^* \approx 7.5$, $Z_{O\|}^* \approx -5.7$, $Z_{O\perp}^* \approx -2.1$. $\mathcal{M} = \frac{1}{4}\sum_k \mathcal{M}_k = 24$ amu is the effective mode mass; the mode kinetic energy is $T_{\mathrm{kinetic}} = \frac{\mathcal{M}\dot{Q}^2}{2}$.

The classical TO mode dynamics can be analyzed using the Landau-Devonoshire free energy[7,9]

$$F = \frac{\mathcal{M}\omega_0^2}{2}|\boldsymbol{Q}|^2 + g_1\boldsymbol{Q}\cdot\boldsymbol{p}_{\mathrm{FE}} + g_2(\boldsymbol{Q}\cdot\boldsymbol{p}_{\mathrm{FE}})^2 - \boldsymbol{E}(t)\cdot eZ^*\boldsymbol{Q}, \tag{S3}$$

where $\omega_0 = 3\times 10^{12}$ s$^{-1}$ is the unperturbed TO mode frequency, and the second and the third terms on the right-hand side account for the two lowest-order symmetry-allowed coupling terms between the *t*-FE polarization and the TO mode[9]. The corresponding equation of motion is

$$\mathcal{M}\ddot{\boldsymbol{Q}} + 2\mathcal{M}\Gamma_{\mathrm{TO}}\dot{\boldsymbol{Q}} + \mathcal{M}\omega_0^2\boldsymbol{Q} + g_1\boldsymbol{p}_{\mathrm{FE}} + 2g_2\boldsymbol{p}_{\mathrm{FE}}(\boldsymbol{Q}\cdot\boldsymbol{p}_{\mathrm{FE}}) = eZ^*\boldsymbol{E}, \tag{S4}$$

where the second term on the left-hand side represents Landau-Khalatnikov relaxation with the rate $\Gamma_{\mathrm{TO}}$. The last two terms on the left-hand side account for the coupling between *t*-FE polarization and the TO mode. The second to last term results in biasing of the TO oscillation, which is experimentally indistinguishable from the *t*-FE mode itself and thus will be hereafter ignored. In contrast, the last term on the left-hand side modifies the oscillation frequency of the TO mode polarization component collinear with $\boldsymbol{p}_{\mathrm{FE}}$,

$$\omega_{||}(\boldsymbol{p}_{FE}) = \sqrt{\omega_0^2 + \frac{2g_2}{\mathcal{M}}|\boldsymbol{p}_{\mathrm{FE}}|^2} \approx (1 + \alpha|\boldsymbol{p}_{\mathrm{FE}}|^2)\omega_0 = \omega_0 + \Delta\omega, \tag{S5}$$

where $\alpha = \frac{g_2}{\omega_0^2\mathcal{M}}$, $\Delta\omega = \frac{g_2|\boldsymbol{p}_{\mathrm{FE}}|^2}{\omega_0\mathcal{M}}$. The polarization-induced TO mode stiffening is described by $\alpha > 0$. This *t*-FE-induced anisotropy is equivalent to the TO phonon splitting into the $A_1$ mode and E modes in a FE state.

This frequency splitting results in beating between two polarization components analogous to the optical birefringence in anisotropic crystals. To describe the effect on ionic magnetization, we select the *x*-axis along $\boldsymbol{p}_{\mathrm{FE}}$, and consider $Q_x = A_x \cos\omega_{||}t$, $Q_y = A_y \cos\omega_0 t$. The *z*-component of corresponding dynamical magnetization $M_z$ is

$$M_z = \frac{eZ^*}{2}\left[\boldsymbol{Q} \times \dot{\boldsymbol{Q}}\right]_z = \frac{eZ^*}{4}A_x A_y[\Delta\omega \sin\Omega t + \Omega \sin\Delta\omega t], \tag{S6}$$

where $\Omega = 2\omega_0 + \Delta\omega$. Since the magnetic contribution to SHG intensity is heterodyned with the polarization-induced contribution, $I_{\mathrm{SHG}}^{\mathrm{CD}} \propto M_z$. Thus, Eq. (S6) describes a superposition of two oscillating contributions to magnetic SHG intensity, consistent with our observations. Moreover, since the *t*-FE polarization decreases with increasing temperature, $\Delta\omega \propto |\boldsymbol{p}_{\mathbf{FE}}|^2$ is expected to decrease, explaining the observed redshift of the slow oscillation. In contrast, $\Omega$ is dominated by $2\omega_0$ does not exhibit a significant variation. Eq. (S6) also predicts that $m$ decreases together with the TO mode oscillation amplitude, which is inconsistent with the observed long decay time of slow oscillation. Next, we develop a minimal quantum model elucidating the origin of this discrepancy.

*2D Quantum model of the QPE state.*

In this subsection, we show that the long-time slow oscillations of magnetic SHG are consistent with the non-classical contribution to ionic magnetization. To analyze quantum ion dynamics driven by elliptically polarized THz pulses, we consider the two-dimensional (2D) ionic displacement energy landscape in the (001) plane. In the one-dimensional (1D) approximation for the QPE state the ionic displacement potential exhibits two symmetric shallow minima in the opposite displacement directions[10]. By the $C_4$ symmetry of the tetragonal phase, these correspond to four minima in the *xy* plane along either the principal or the <11> directions.

Scanning transmission electron microscopy (STEM) studies of a sister compound $BaTiO_3$ characterized by deeper displacement potential wells show eight displacement energy minima along the <111> family of directions[11], which persist even in the cubic paraelectric phase. A similar energy landscape with minima along the eight <111> directions in STO is suggested by diffuse X-ray scattering[12], which for concreteness will be assumed in the analysis below.

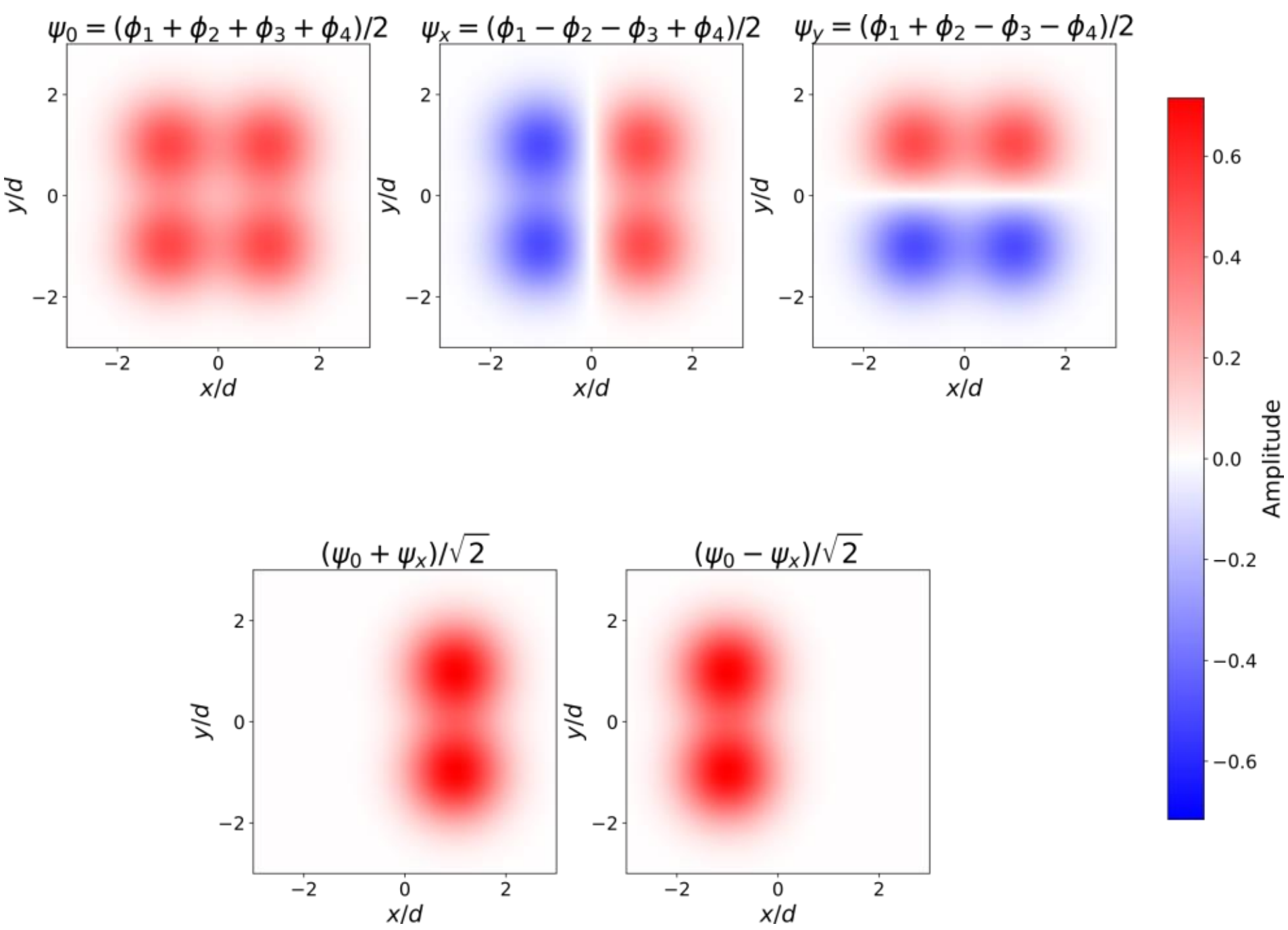

**Figure S1. (Top) Pseudo-color maps of the stationary wavefunctions of the tight-binding Hamiltonian. (Bottom) Illustration of superpositions of the stationary states describing polarization along the *x*-axis.** The basis wavefunctions are approximated as superpositions of Gaussians with half-width $\sigma = 2d/3$ where the local potential mimima are located at $(\pm d, \pm d)$.

However, with appropriately defined basis wavefunctions our analysis remains valid if the minima are located along the principal directions. For displacements in the *xy* plane, and in the absence of *t*-FE polarization, the minima project onto four symmetric minima positioned at ($x$, $y$) = ($\pm d$, $\pm d$).

We use the tight-binding approximation to analyze the lowest-energy ionic states, in the tight binding basis of the wavefunctions $\phi_0$, $\phi_1$, $\phi_2$, and $\phi_3$ quasi-localized in the minima at the positions $(d, d)$, $(-d, d)$, $(-d, -d)$ and $(d, -d)$, respectively. In the absence of *t*-FE, hybridization (hopping amplitude) between the neighboring minima is given by

$$\langle\phi_j|H|\phi_{j\pm1}\rangle = -V/2. \tag{S7}$$

The ground state $\psi_0 = \frac{1}{2}\sum_{j=0}^{3}\phi_j$ of this Hamiltonian has energy $E_0 = -V$. Here, we neglect the corrections to normalization due to the non-orthogonality of the basis wavefunctions. The two lowest-energy excited states, $\psi_x = \frac{1}{2}(\phi_0 - \phi_1 - \phi_2 + \phi_3)$, $\psi_y = \frac{1}{2}(\phi_0 + \phi_1 - \phi_2 - \phi_3)$ are degenerate with energy $E_1 = 0$. The spatial profiles of these states are illustrated in Fig. S1 (top).

The dipole moment $\hat{\boldsymbol{p}} = eZ^*\hat{\boldsymbol{r}}$ has finite matrix elements between the ground state and the excited states,

$$\langle\psi_x|\hat{\boldsymbol{p}}|\psi_0\rangle = (p_0, 0);\ \ \langle\psi_y|\hat{\boldsymbol{p}}|\psi_0\rangle = (0, p_0), \tag{S8}$$

where $p_0 = eZ^*d$. Accordingly, polarization along the *x*-axis is described by superpositions of $\psi_0$ and $\psi_x$ (Fig. S1, bottom row), and polarization along the *y*-axis – by superpositions of $\psi_0$ with $\psi_y$. The value $V = 2$ meV is evaluated by matching the frequency of the TO mode to the splitting between the ground and the first excited states, $\hbar\omega_{\mathrm{TO}} = V$. The equilibrium populations of the excited states $n_i = e^{-E_i/kT}/Z$ , where $Z = \sum_i e^{-E_i/k_BT}$ is the partition function, are $n_x = n_y = 0.08$ at $T = 10$ K, and are an order of magnitude smaller than $n_0 = 0.84$. The population of the next excited state with energy $2V$ relative to the ground state is negligible, indicating that THz-driven polarization dynamics can be well-approximated by utilizing the basis of three lowest-energy wavefunctions $\psi_0$, $\psi_x$ and $\psi_y$. Hereafter, we refer to them as the basis states.

The magnetic moment operator $\widehat{\boldsymbol{M}} = \frac{eZ^*}{2\mathcal{M}}\hat{L}$, where $\hat{L} = -i\hbar(x\partial_y - y\partial_x)$ is the angular momentum operator, has finite matrix elements only between the exited states

$$\langle\psi_y|\widehat{\boldsymbol{M}}|\psi_x\rangle = \frac{i\hbar eZ^*C}{2\mathcal{M}} \equiv iM_0. \tag{S9}$$

Here $C = \langle\psi_y|y\partial_x - x\partial_y|\psi_x\rangle$ is a constant of order 1 determined by the spatial profiles of the ionic wavefunctions. Matching with the classical value $M_c = \frac{\boldsymbol{p}\times\dot{\boldsymbol{p}}}{2eZ^*}$ for small amplitudes of $\psi_x$ and $\psi_y$ in the trial wavefunction $\psi = \alpha\psi_0 + e^{-i\omega_{\mathrm{TO}}t}\beta(\psi_x \pm i\psi_y)$ yields $C = \frac{d^2}{l^2}$, where $l = \sqrt{\frac{\hbar}{\mathcal{M}\omega_{\mathrm{TO}}}}$ is the oscillator length.

Magnetic moment is carried by complex superpositions of $\psi_x$ and $\psi_y$, and is maximized in the states $\psi_{L,R} = e^{\mp i\pi/4}(\psi_x \pm i\psi_y)/\sqrt{2}$, $\langle\psi_{L,R}|\widehat{\boldsymbol{M}}|\psi_{L,R}\rangle = \pm M_0$ (Fig. S2). This analysis demonstrates that quantum ionic states can exhibit magnetization even in the absence of dynamical polarization. The dipolar moment is carried only by the coherent superpositions of the ground state with $\psi_x$ and $\psi_y$, while the magnetization only requires coherent superpositions of $\psi_x$ and $\psi_y$. In particular, the states $\psi_{L,R}$ carry the maximum magnetic moment but no dipole moment. On the other hand, the classical circularly polarized displacement states have the form

$$\psi^{C}{}_{L,R} = \frac{1}{\sqrt{2}}\psi_0 + \frac{e^{\mp\frac{i\pi}{4}}\psi_x + e^{\pm i\pi/4}\psi_y}{2} = \frac{1}{\sqrt{2}}(\psi_0 + \psi_{L,R}), \tag{S10}$$

carrying only a much smaller magnetization than the full magnetization of $\psi_{L,R}$.

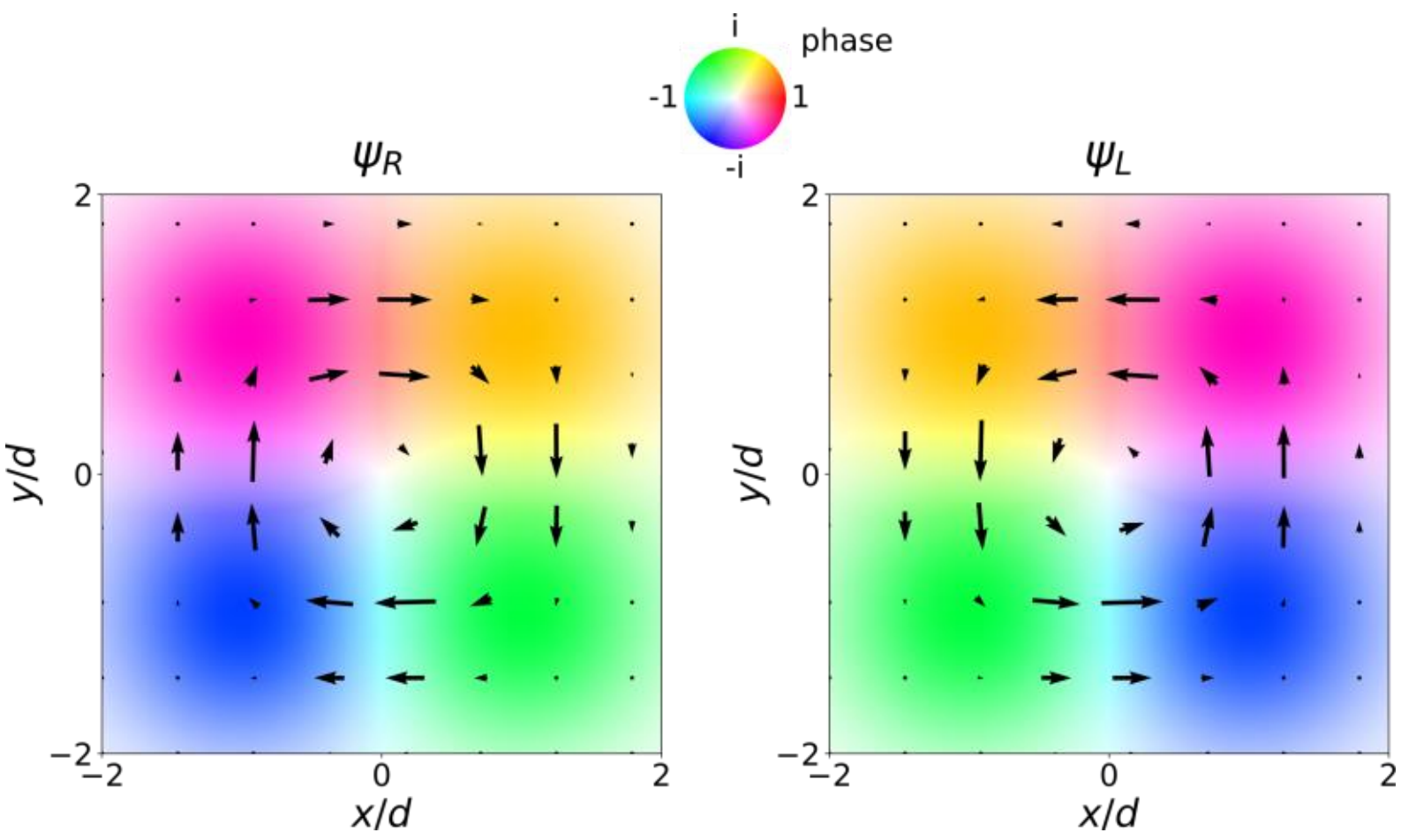

**Figure S2. Pseudo-color maps of the magnetic moment-carrying states, as defined in the text, in the same approximation as in Fig. S1.** Arrows show the current density $J = \frac{\hbar}{\mathcal{M}}\mathrm{Im}(\psi^*\nabla\psi)$.

In the presented model, the nonlinear frequency shift due to *t*-FE polarization is a consequence of the modification of the corresponding single-ion hopping amplitudes, $V_x = V + \Delta V$ for *t*-FE polarization, with $\Delta V = \alpha|\boldsymbol{p}_{\mathrm{FE}}|^2 V$. The resulting energy of the ground state is $E_0 = -V - \Delta V/2$, while the energies of the excited states are $E_x = \Delta V/2$, $E_y = -\Delta V/2$. The frequency of the *y*-polarized TO mode remains unchanged, while the frequency of the *x*-polarized mode is increased by $\Delta\omega = \Delta V/\hbar$. For the tetragonal domains with in-plane orientation of the *c*-axis, similar splitting is also expected due to the anisotropy of the TO phonon dispersion, even in the absence of t-FE polarization. Such domains are equally split between *c*-axis orientations along the two in-plane principal directions, cancelling their contributions to the magnetic SHG signal. A similar cancellation is expected for the mode splitting due to randomly oriented polar nanodomains.

*Quantum model of ionic dynamics.*

In the basis of states $\psi_0$, $\psi_x$, $\psi_y$, the ionic Hamiltonian in the presence of the THz field $\boldsymbol{E}(t) = (E_x(t), E_y(t))$ is

$$H = \begin{bmatrix} -\hbar\omega_0 & -p_0 E_x(t) & -p_0 E_y(t) \\ -p_0 E_x(t) & \alpha\hbar\omega_0 p_{\mathrm{FE},x}^2/2 & 0 \\ -p_0 E_y(t) & 0 & \alpha\hbar\omega_0 p_{\mathrm{FE},y}^2/2 \end{bmatrix}. \tag{S11}$$

We approximated the experimentally measured THz pump pulse by a sinusoidal function convolved with a Gaussian, $E_{x,y} = E_{0,x,y}\cos(\Omega t - \varphi_{x,y})e^{-\left(\frac{t\Omega}{\sigma}\right)^2/2}$, with the parameters $E_{0,x} = \pm 0.7 E_{0,y}$ for RCP/LCP pulses, $\Omega = 3.3$ ps$^{-1}$, $\varphi_x = -0.25$, $\varphi_y = 1.37$, $\sigma = 2.1$ determined by fitting the results of electro-optic sampling. We use $\tau_{\mathrm{FE}} = 20$ ps to model the observed decay of $t$-FE polarization. The TO mode nonlinearity coefficient α is similarly estimated based on the observed magnetic beating frequency observed at 10 K, and the relaxation parameters $\gamma_{ph} = 0.5$, $\gamma_E = 0.2$ are used. The temperature dependence of the TO mode frequency is approximated by the Taylor expansion of the simplified Barrett formula[13] $\omega_{TO}^2(T) = \omega_0^2 + \eta T^2$, where $\eta = 2.0 \times 10^{-3}$ $\mathrm{THz}^2/K^2$ and the $t$-FE mode relaxation rate by $\Gamma(T) = \Gamma(0)e^{-T/T_{QPE}}$ phenomenologically accounting for the suppression of $t$-FE above the transition at $T_{QPE} = 40$ K.

The quantum ionic dynamics are analyzed using the Lindblad master equation

$$\frac{d\rho}{dt} = -\frac{i}{\hbar}[H, \rho] + D[\rho], \tag{S12}$$

where $\rho = |\psi_i\rangle\langle\psi_i|$ is the density matrix in the basis of $\psi_0$, $\psi_x$, $\psi_y$ and $D[\rho]$ is the dissipation function. Since $t$-FE relaxation is likely spatially inhomogeneous, we expect that relaxation of TO mode is dominated by dephasing due to the non-linear $t$-FE-TO coupling, which is approximated by $D_{\mathrm{ph}}[\rho]_{ij} = -\frac{\gamma_{ph}}{\hbar}|H_{ii} - H_{jj}|\rho_{ij}$, with dimensionless dephasing parameter $\gamma_{ph}$ given by the rate of dephasing between the stationary states scaled by their frequency difference. In addition, the Landau-Khalatnikov relaxation towards equilibrium is described approximately by $D_{\mathrm{E}}[\rho] = -\gamma_{\mathrm{E}}(\rho - \rho_0)$, such that $D = D_{\mathrm{E}} + D_{\mathrm{ph}}$. In the simulations, we used empirical values $\gamma_{\mathrm{ph}} = 0.4$ ps$^{-1}$ and $\gamma_{\mathrm{E}} =$ 0.2 ps$^{-1}$.

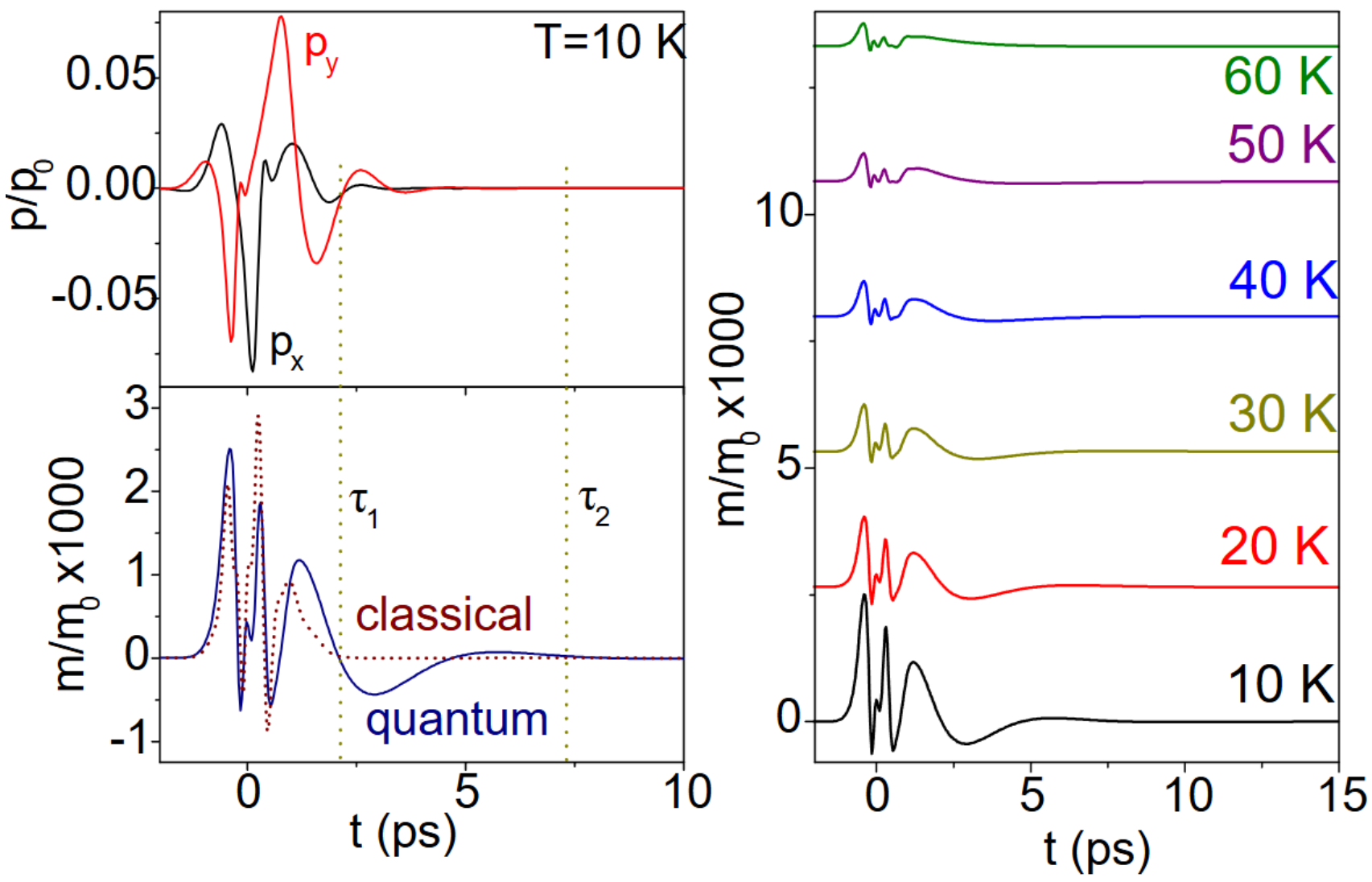

**Figure S3. Left top: Time traces of dipole moment components $p$ calculated for $T = 10$ K. Left bottom: Classical vs quantum $M(t)$, at $T = 10$ K.** Vertical dotted lines show the times $t_1$ ($t_2$) when classical (quantum) magnetization becomes negligible. Right: time traces of quantum $M(t)$, calculated at the specified temperatures.

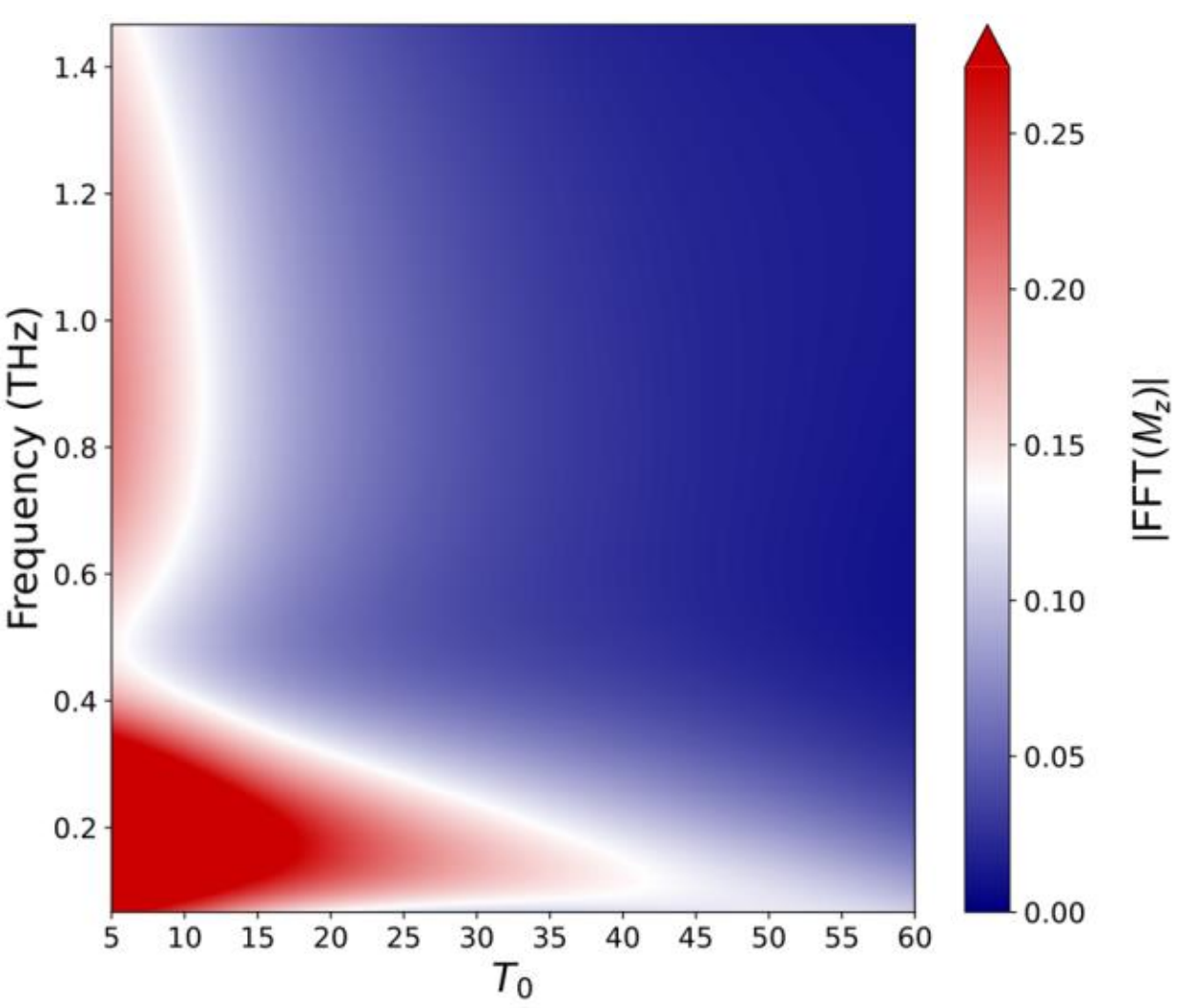

**Figure S4**. **Pseudo-color map of the Fourier transform of quantum *M*(*t*).**

The elements of the density matrix have a simple physical interpretation: the diagonal elements are the populations of the corresponding basis states, while the off-diagonal elements describe coherences that carry electric dipole and magnetic moments: $\mathrm{Re}(\rho_{01}) = p_x/2p_0$, $\mathrm{Re}(\rho_{02}) = p_y/2p_0$, $\mathrm{Im}(\rho_{12}) = M/2M_0$. The electric dipole and the (quantum) ionic magnetic moments are determined by different coherences, allowing the ionic magnetic moment to persist even after the oscillating polarization decays.

These qualitative insights are supported by simulations of the dynamical ionic magnetization (Figs. S3 and S4), which reproduce the essential features of the measured magnetic SHG, including the rapid short-time oscillations and the slow oscillations at longer times whose amplitude and frequency decrease with increasing temperature. The decrease of amplitude is a consequence of increasing thermal occupation of the excited states, resulting in reduced coherence amplitudes driven by their THz-induced excitation from the ground state. Meanwhile, the decrease of frequency results from the reduced nonlinear frequency shift produced by *t*-FE polarization, which becomes strongly suppressed at higher temperatures due to enhanced thermal fluctuations and the concomitant disappearance of polar nanoregions that may help stabilize the t-FE state.

In the simulations, the oscillation around 1 THz is notably broader, resulting in a less pronounced peak in the frequency spectrum (Fig. S4). This discrepancy likely originates from the leakage of the polarization-induced SHG signal of non-magnetic origin into the measured SHG-CD.

## Section S1. Characterization of circularly polarized THz pulse

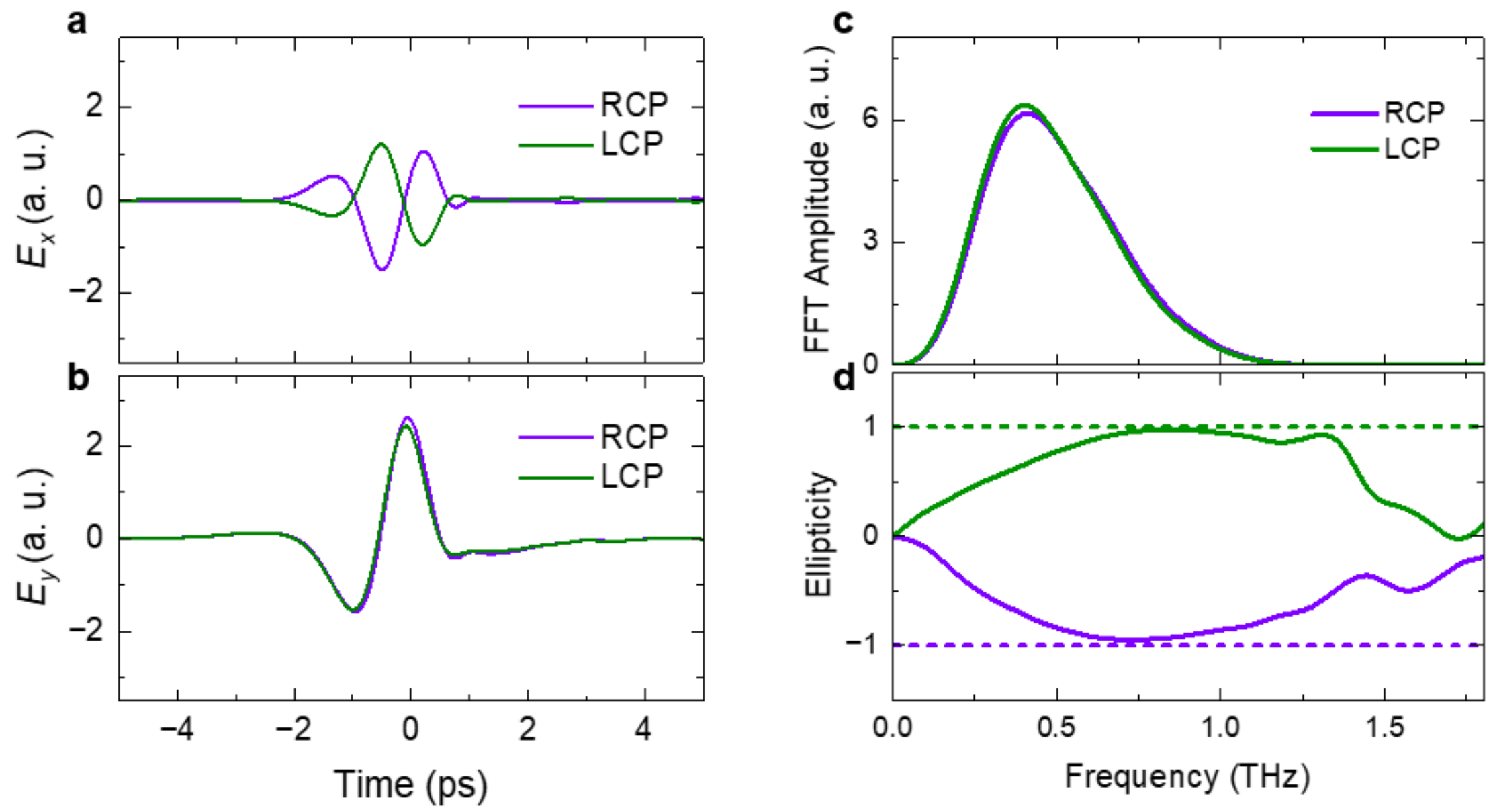

**Figure S5. Time profile of elliptically polarized THz pulse obtained through electro-optic sampling.**

To explore THz-pump-induced quantum ionic magnetization in STO, we induced elliptically polarized single-cycle THz pulses by introducing a time delay between two orthogonal THz pulses with right-handed (RCP, purple) and left-handed (LCP, green) helicities. Figures S5a and S5b show the time profiles of $E_x$ and $E_y$ components of the elliptically polarized THz pulses, respectively. When the helicity of THz pulse is reversed, only the $E_x$ component flips sign without changing the waveform, while the $E_y$ component remains unchanged. This nearly symmetric waveform between RCP and LCP THz pulses helps minimize artifacts emerging from asymmetry. We note that the field amplitude of the $E_x$ component is two times smaller than that of the $E_y$ component, resulting in the elliptical THz pulse. Figures S5c and S5d show FFT spectra for RCP and LCP THz pulses, and ellipticity spectra calculated from the Stokes parameter[14], $\eta = \frac{2\mathrm{Im}(E_x E_y^*)}{|E_x|^2+|E_y|^2}$. Both RCP and LCP THz pulses show identical amplitudes in the frequency domain, with their ellipticity reversed, confirming opposite helicity. At the frequency of 0.75 THz, the ellipticity of RCP and LCP THz pulses is –1 and +1, respectively, which are perfectly circular. This facilitates effective excitation of axial transverse optical (TO) phonons of similar frequency in STO[15].

## Section S2. THz-pump-induced second harmonic generation with linear probe polarization

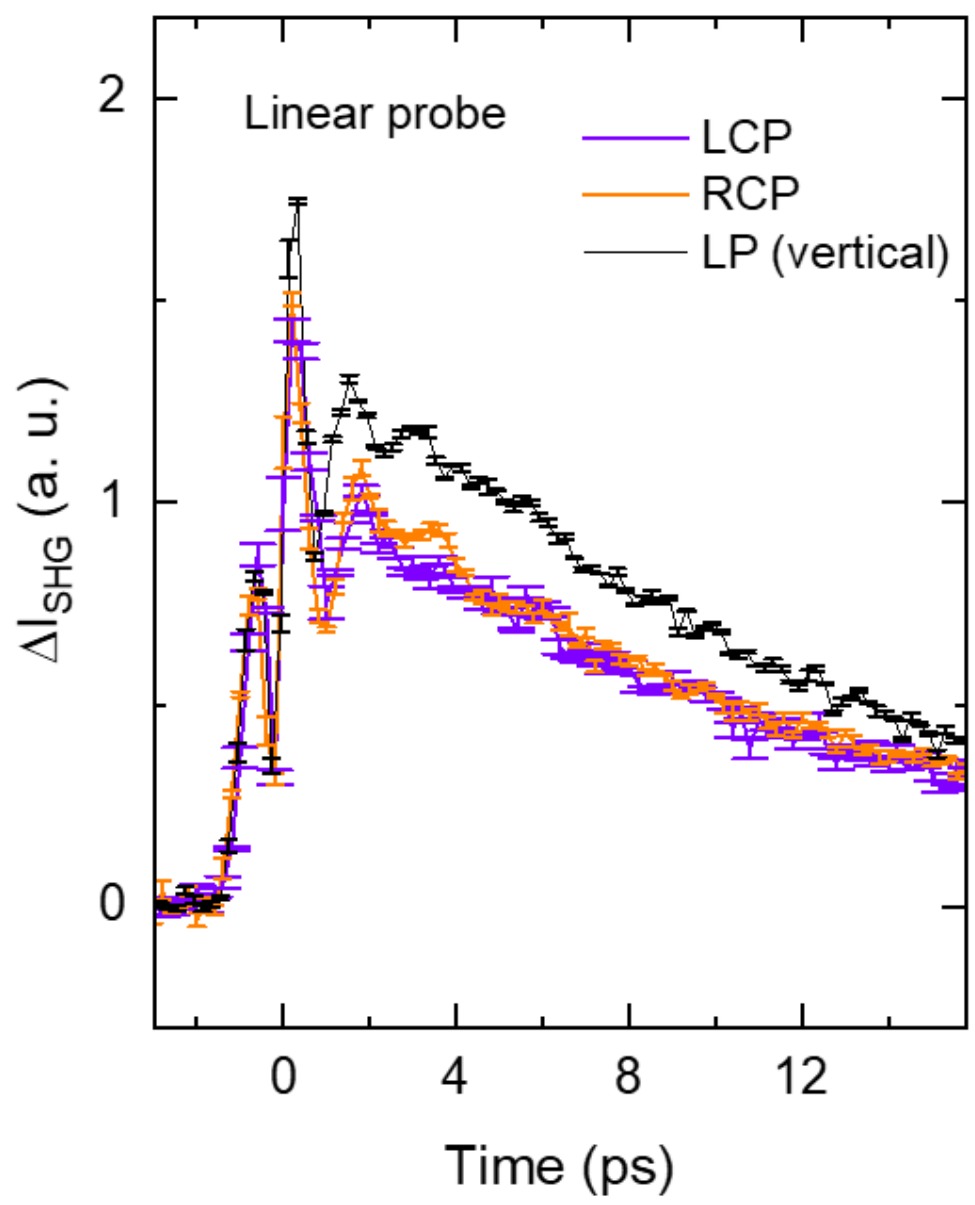

**Figure S6. THz-pump-induced SHG signals $\Delta I_{SHG}$ for left-handed (LCP), right-handed (RCP), linear (LP) THz pulse obtained with linear probe polarization.**

To confirm THz-pump-induced inversion symmetry breaking under elliptically polarized THz excitation, we obtained transient SHG signals for left-handed (LCP, purple), right-handed (RCP, orange), and linear (LP, black) THz pulses with linear probe polarization at $T = 10$ K (Fig. S6). By using an analyzer, we only monitored vertically-polarized SHG signals parallel to the LP THz pulse and the [010] crystal orientation of STO. All THz polarization configurations show similar dynamics without slow oscillations, which only appear when using a circular optical probe that is sensitive to time-reversal–symmetry breaking. The finite non-oscillatory SHG signal observed under elliptically polarized THz excitation indicates inversion-symmetry breaking. The absence of slow oscillations is also consistent with our theoretical expectation (Supplementary Note 1). The primary difference between elliptically and linearly polarized THz pulses is their overall intensity, which is reduced due to the diminished vertical *E*-field component introduced during the conversion from linear to elliptical polarization.

**Section S3. Comparison between second harmonic generation circular dichroism under linearly and circularly polarized THz pumping**

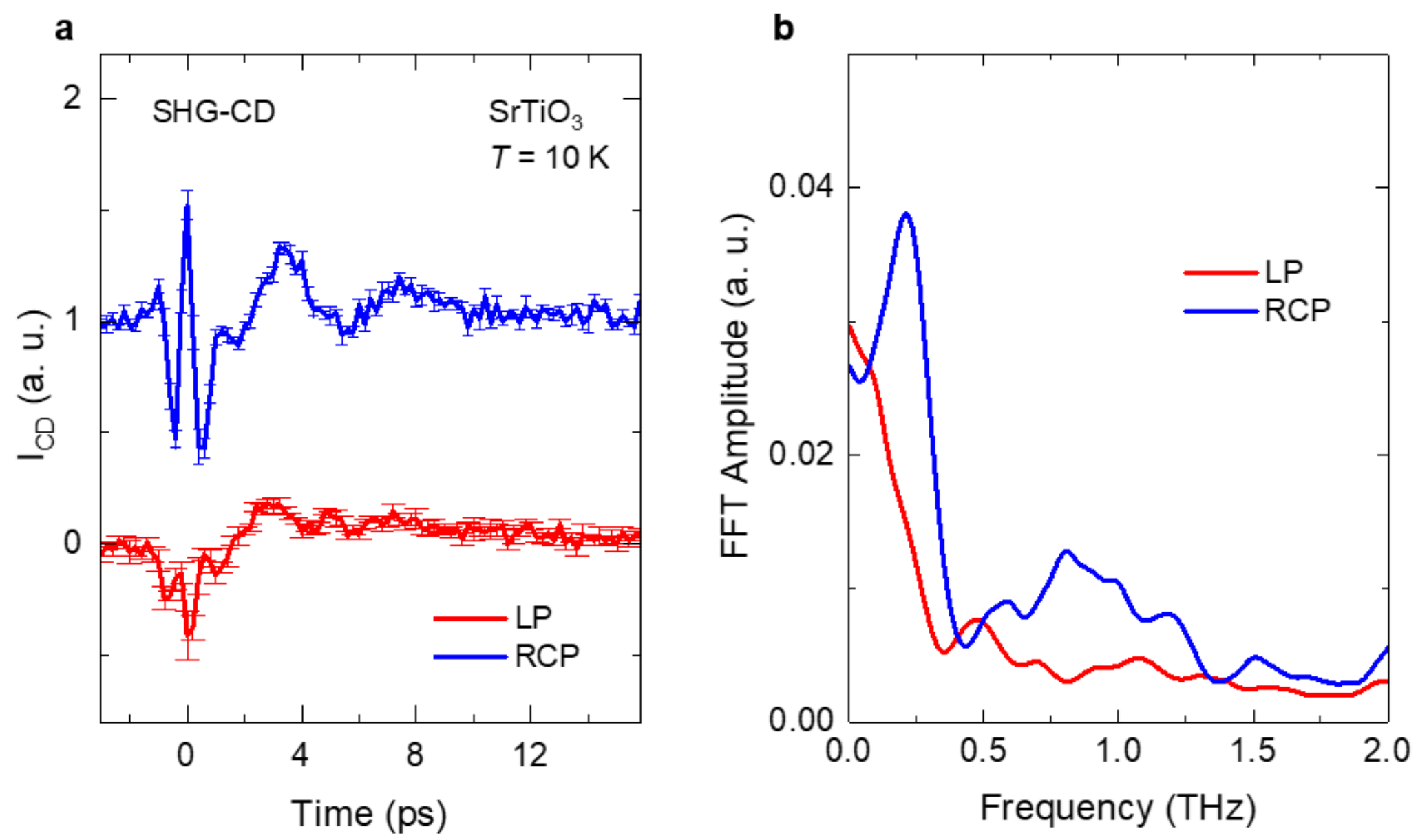

**Figure S7. Linear (LP), and right-handed (RCP) THz-pump-induced SHG circular dichroism $I_{CD}$ obtained at $T$ = 10 K.**

Oscillatory signals observed in $I_{CD}$ exhibit a THz helicity-dependence, indicating time-reversal symmetry breaking. On the other hand, we can expect the absence of both fast and slow oscillations in $I_{CD}$ under linearly polarized THz excitation. Figure S7a shows THz pump-induced SHG-CD obtained with linear (LP, red), and right-handed (RCP, blue) THz pulses. Notably, both fast and slow oscillations which can be observed in RCP THz pumping disappear under LP THz excitation. The FFT spectra further highlight the absence of the oscillations for LP THz pumping, as shown in Figure S7b. We note that both LP and RCP THz-field–induced $I_{CD}$ signals contain sizable DC components, which can be attributed to distortions of the THz pump and optical probe polarization states due to the thick TPX cryostat window.

## Section S4. THz-helicity-independent second harmonic generation circular dichroism

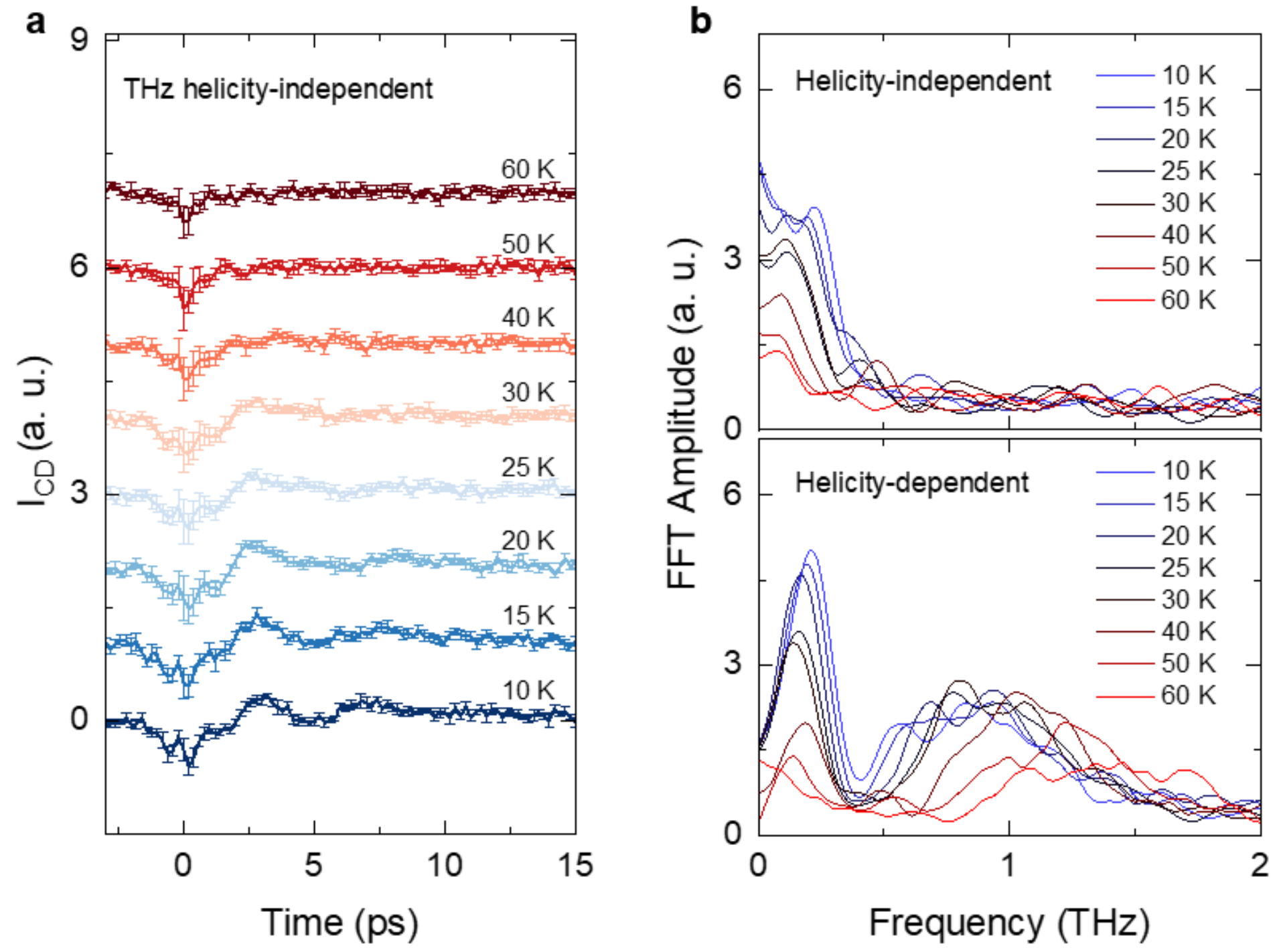

**Figure S8. THz-helicity-independent SHG circular dichroism $I_{CD}$ obtained at several temperatures.**

The asymmetry between the two THz helicities and slight sample misalignment can lead to unintended THz-helicity-independent signals. Figure S8a displays THz-helicity-independent $I_{CD}$ contributions obtained by adding $I_{CD}$(RCP) + $I_{CD}$(LCP) and at several temperatures. At $T$ = 10 K, $I_{CD}$ exhibits a weak slow oscillation and the absence of a fast oscillatory signal, and the slow oscillation disappears as temperature increases. Notably, all THz-helicity-independent $I_{CD}$ signals exhibit a negative, non-oscillatory envelope around $t$ = 0 ps, similar to the $I_{CD}$ response observed under linearly polarized THz excitation. These significant DC contributions are more clearly highlighted in FFT spectra, as shown in Fig. S8b. In the spectra of the THz-helicity-independent $I_{CD}$ (top panel), exponentially decaying DC components dominate similar to $I_{CD}$ under linear THz pumping, whereas the THz-helicity-dependent spectra (bottom panel) show clear peak-like features. This suggests that the THz-helicity-independent contributions result from residual linear polarization components of the THz field, which can be cancelled in $I_{CD}$(RCP) – $I_{CD}$(LCP).

## Section S5. THz helicity-dependent linear dichroism

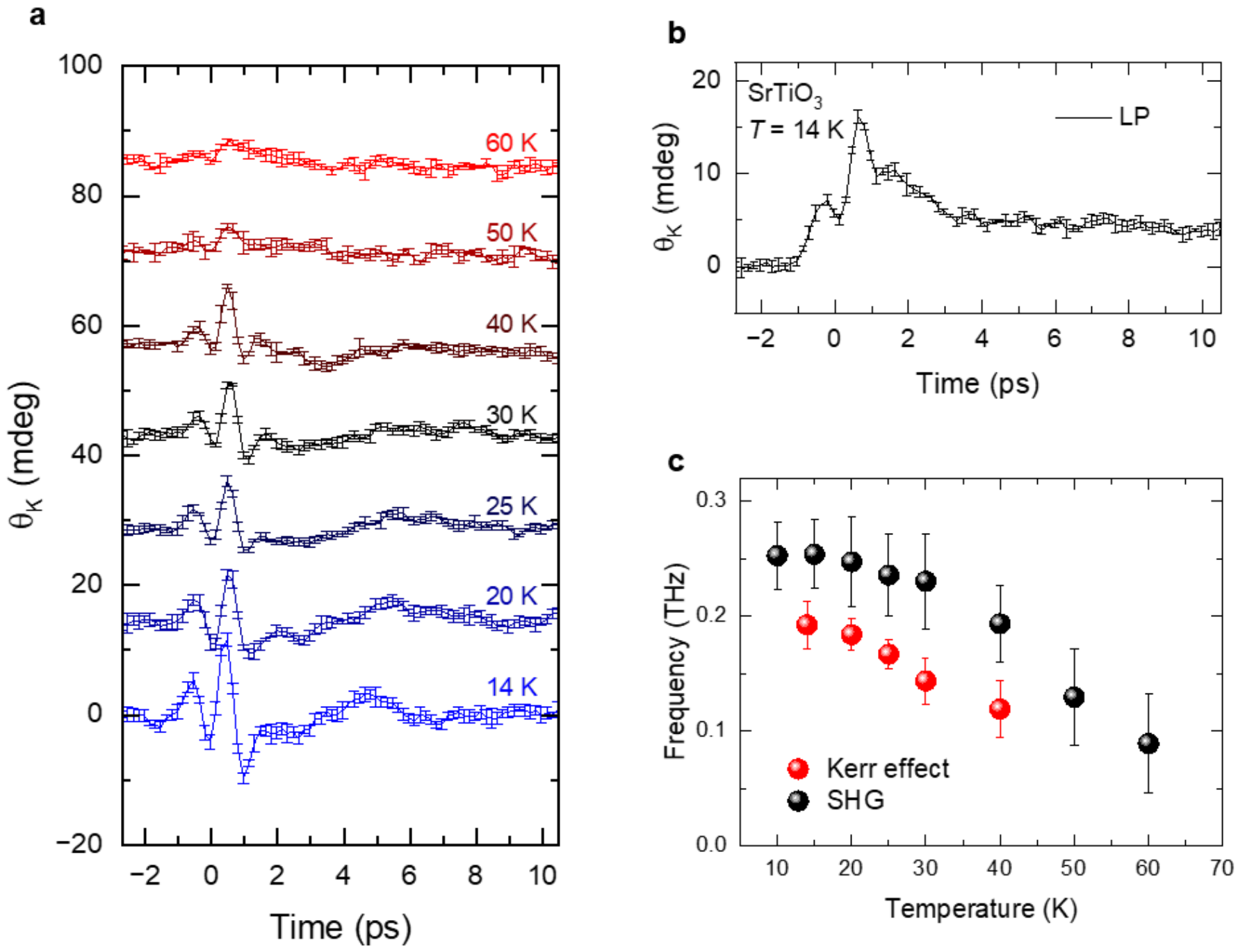

**Figure S9. THz-pump-induced linear dichroism measurements under RCP and LCP THz excitation.**

THz field-induced ionic magnetization was also observed in linear dichroism (LD) measurements, exhibiting a non-zero THz helicity-dependent response. Under LP THz excitation, THz-pump-induced polarization anisotropically modulates the refractive index of STO, leading to polarization rotation $\theta_K$ with fast oscillatory signals[16,17]. On the other hand, an oscillating magnetic field can further modulate the refractive index as described by the time non-invariant axial magneto-optic tensor $q_{ijk}$[18]. The magnetic-field-induced dielectric constant change $\Delta\varepsilon_{ij}$ is simply proportional to the magnetic field $B$ and can be written as $\Delta\varepsilon_{ij} = q_{ijk}B_k$. In the magnetic point group $m'$, there are 10 non-zero $q_{ijk}$ components, adapted from Bilbao crystallographic server[19], as indicated in Table S3. This allows to the out-of-plane $B_z$-dependent non-zero LD signals $\Delta\varepsilon_{xx}$–$\Delta\varepsilon_{yy}$ = ($q_{xxz}$–$q_{yyz}$)$B_z$, enabling us to observe THz-pump-induced magnetic oscillations in STO.

**Table S3. Non-zero tensor components of $q_{ijk}$, and modulated dielectric constants under the magnetic field $B$. Contributions from the out-of-plane components $B_z$ are highlighted in red.**

| | Time non-invariant axial magneto-optic tensor $q_{ijk}$ |
|---|---|
| Non-zero components | $q_{xxx}$, $q_{xxz}$, $q_{yyx}$, $q_{yyz}$, $q_{zzx}$, $q_{zzz}$, $q_{yxy}$, $q_{xzx}$, $q_{xzz}$, $q_{xyx}$, |
| Dielectric constant $\boldsymbol{\Delta\varepsilon_{ij}}$ | $\Delta\varepsilon_{ij} = \begin{pmatrix} q_{xxx}B_x + q_{xxz}B_z & q_{xyx}B_x + q_{yxy}B_y & q_{xzx}B_x + q_{xzz}B_z \\ q_{xyx}B_x + q_{yxy}B_y & q_{yyx}B_x + q_{yyz}B_z & 0 \\ q_{xzx}B_x + q_{xzz}B_z & 0 & q_{zzx}B_x + q_{zzz}B_z \end{pmatrix}$ |

Figure S9a displays THz-helicity-dependent $\theta_K$ obtained at several temperatures. We set the input polarization to be 45 degrees with respect to the [010] direction of STO, and measured $\theta_K$ through balance detection. We observed both fast and slow oscillatory signals in THz-helicity-dependent $\theta_K$, consistent with SHG-CD, whereas under linearly polarized (LP) THz excitation only the fast oscillatory signal appears (Fig. S9b), as previously reported[16]. The observation of both fast and slow oscillatory signals only in THz helicity-dependent $\theta_K$ further supports the emergence of magnetic oscillations in STO under elliptically polarized THz excitation. Furthermore, the frequency of the slow oscillatory signals is slightly lower than that observed in SHG-CD, but the overall temperature-dependent trend is similar, suggesting a common origin (Fig. S9c).

## Section S6. Magnetic and non-magnetic contributions in second harmonic generation circular dichroism

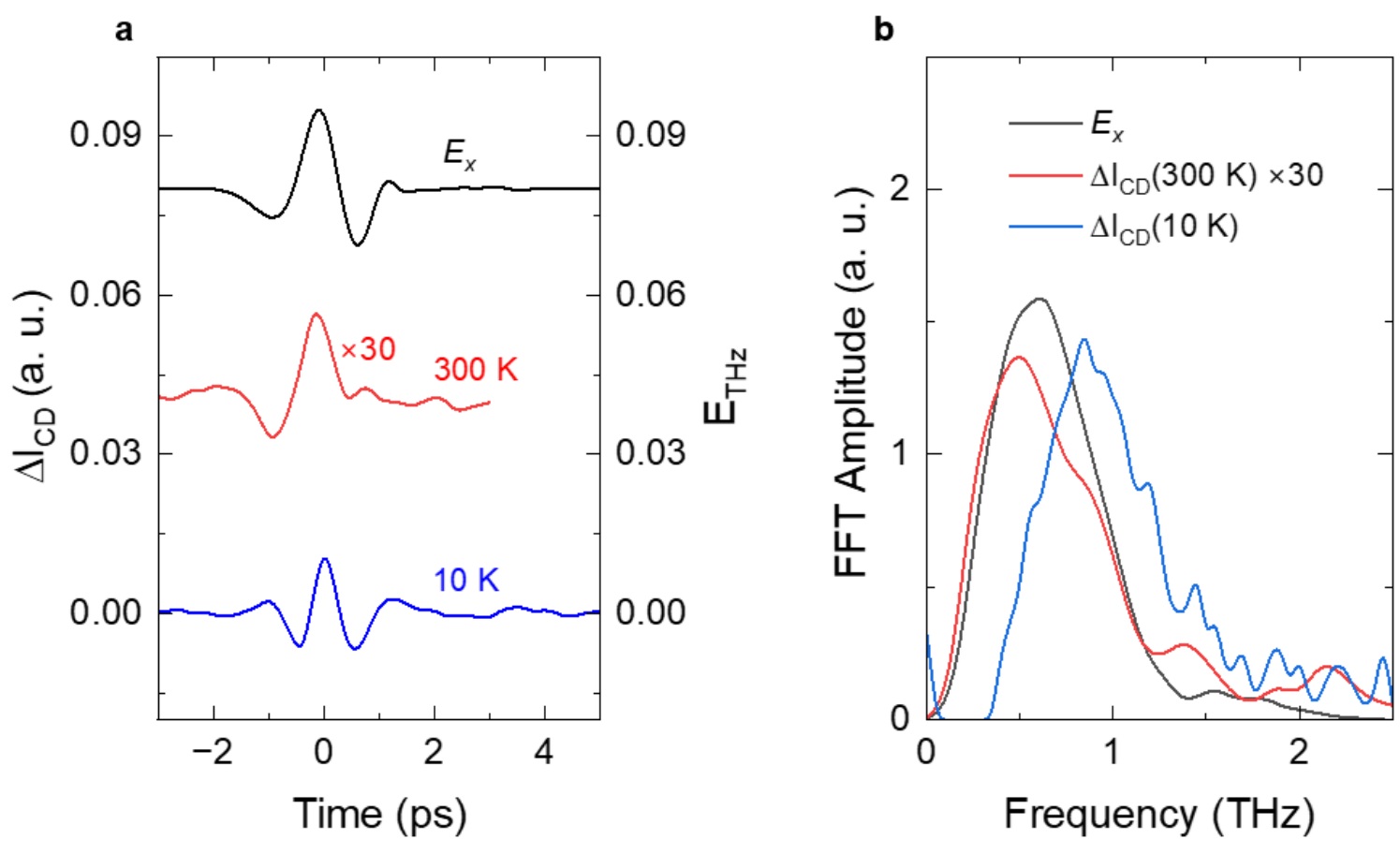

**Figure S10. THz-helicity-dependent SHG circular dichroism $\Delta I_{CD}$ obtained at 10 K and 300 K, and the *x*-component of THz field $E_x$.**

THz-field-induced SHG (TFISH) can generate second-harmonic light through a third-order process, where the transient SHG signals follow a THz pulse. In particular, TFISH in STO can exhibit a non-zero THz-helicity-dependent TFISH-CD signal due to a finite background SHG signal[4,5] . Based on our expectation (Supplementary Note 2), the non-magnetic TFISH signals should be proportional to the $E_x$ component of the THz field in both cubic and tetragonal phases[20]. Figure S10a displays THz-helicity-dependent SHG-CD signals $\Delta I_{SHG}$ obtained at 10 K (blue) and 300 K (red), and $E_x$ (black) obtained from electro-optic sampling. We note that the signal amplitude at 300 K was scaled up by a factor of 30 compared to that at 10 K. At room temperature, $\Delta I_{SHG}$ exhibits oscillatory behavior with alternating negative and positive values, reflecting the heterodyne contribution from the background SHG signal, consistent with our expectations. On the other hand, $\Delta I_{SHG}$ obtained at 10 K also exhibits oscillatory behavior, but the waveform is significantly different from $E_x$. We compared the FFT spectra of them, as shown in Fig. S10b. It is noteworthy that the center frequency of $\Delta I_{SHG}$ at 10 K is around 1 THz, while that at 300 K exhibits a peak position similar to that of $E_x$. This strongly suggests that the fast oscillatory signals at low temperature cannot be explained solely by non-magnetic TFISH, for which the signal is expected to be simply proportional to $E_x$. Instead, they likely involve another contribution, such as the ionic effect we anticipated. Furthermore, the fast oscillatory signals decrease rapidly with increasing temperature and vanish above the quantum paraelectric transition temperature (~40 K), supporting their origin from the quantum ionic effect (Fig. S11).

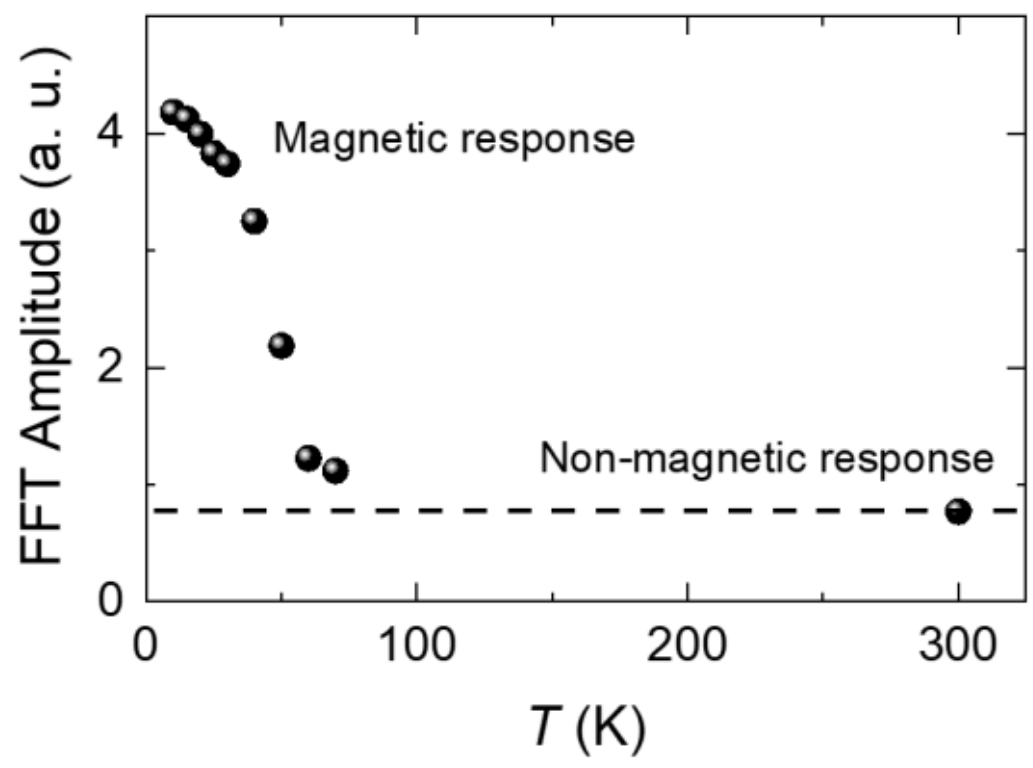

**Figure S11. FFT amplitude of fast oscillatory signals as a function of temperature.**

**Section S7. Temperature-dependent THz-pump-induced second harmonic generation for linear probe polarization**

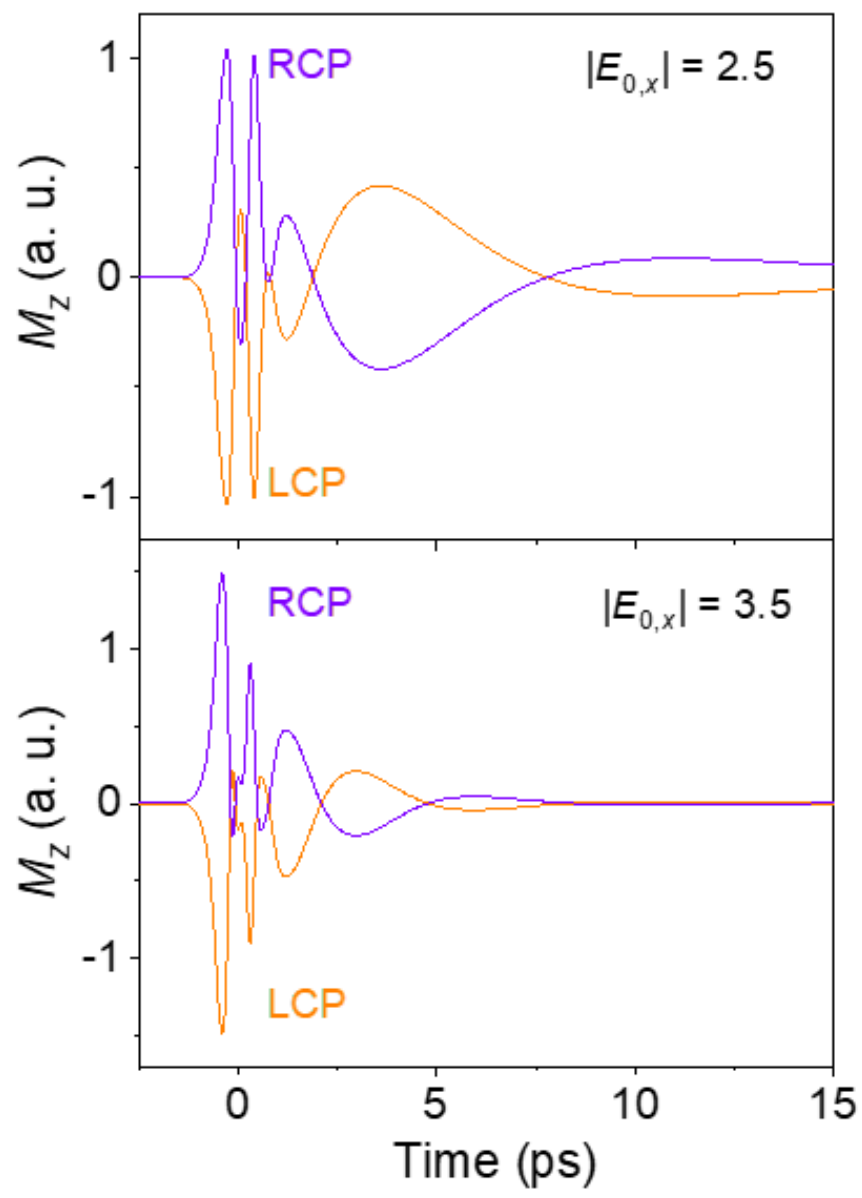

**Figure S12. Numerical simulation of quantum ionic magnetization $M_z$ under RCP and LCP THz excitation obtained at normalized THz $E$-field amplitude $|E|$ = 2.5 (top) and 3.5 (bottom).**

Figure S12 displays the simulated $M_z$ obtained under RCP and LCP THz pumping at different $E$-field amplitude $|E|$. Both fast and slow oscillations in $M_z$ clearly exhibit THz helicity-dependence with opposite sign, implying the origin from the elliptical polarization of the THz pulse. Furthermore, the frequency of the slow oscillation is significantly softened as $|E|$ decreases due to the reduction of $\omega_x - \omega_y$, depending on the amplitude of the non-oscillatory polarization.

## Section S8. Temperature-dependent THz-pump-induced second harmonic generation for linear probe polarization

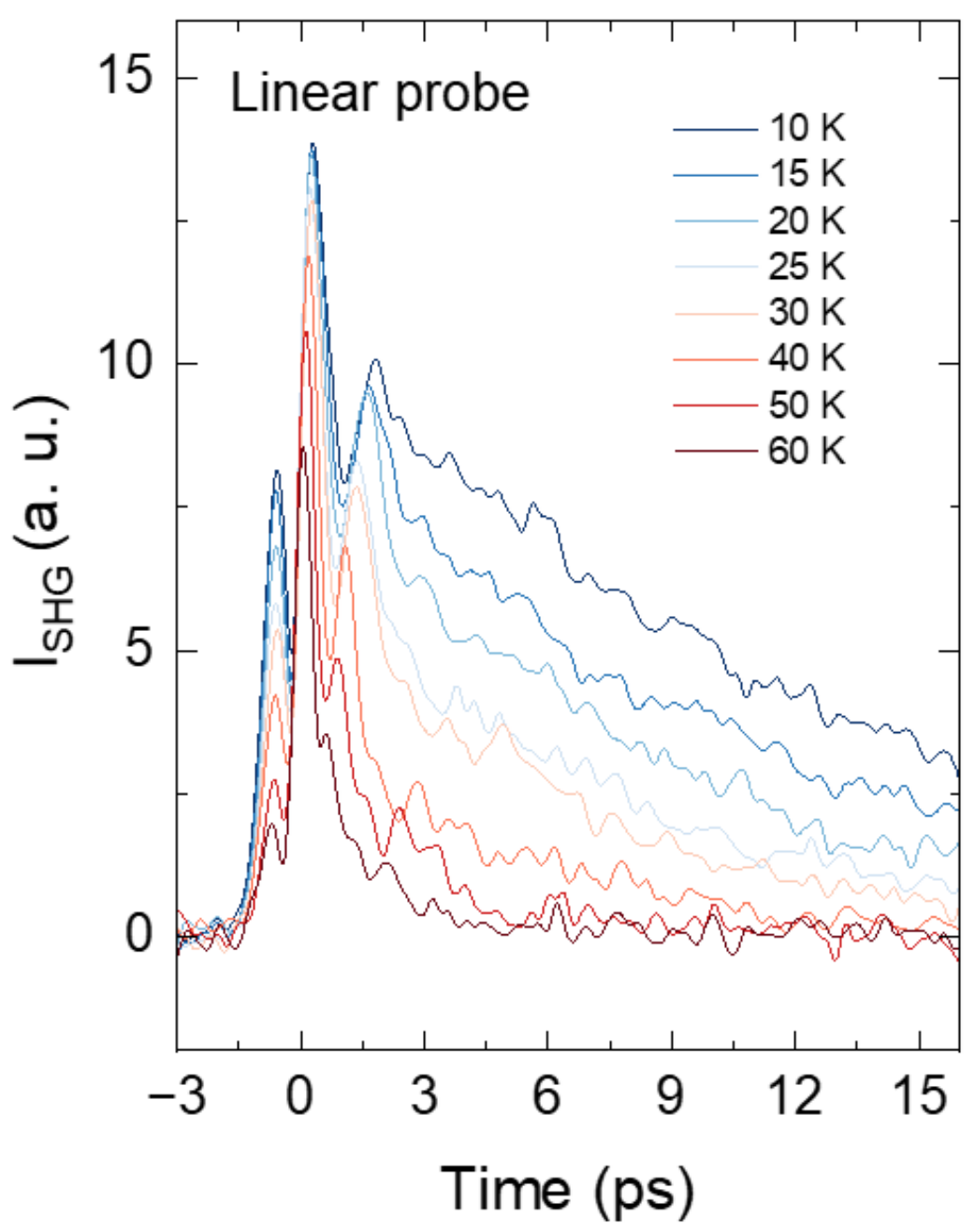

**Figure S13. Temperature-dependent THz-induced SHG $I_{SHG}$ obtained with linear probe polarization.**

Based on our theoretical model, the frequency of slow oscillatory signals observed in SHG-CD is proportional to the THz-induced non-oscillatory polarization amplitude (Supplementary Note 3). Figure S13 shows THz-induced SHG signals $I_{SHG}$ obtained with linearly polarized optical probe and THz pump at several temperatures. The non-oscillatory signals appearing after THz excitation reach their maximum at 10 K and rapidly decrease with increasing temperature. This suggests that the THz-induced non-oscillatory polarization is reduced as temperature increases, contributing to the frequency softening of the slow oscillatory signals in SHG-CD upon heating.

## References

1. Aschauer, U., and Spaldin, N. A. Competition and cooperation between antiferrodistortive and ferroelectric instabilities in the model perovskite $SrTiO_3$. *Journal of Physics: Condensed Matter* **26**, 122203 (2014).
2. Birss, R. R., Symmetry and magnetism. (North-Holland, 1964).
3. Pershan, P. Nonlinear optical properties of solids: energy considerations. *Physical Review* **130**, 919 (1963).
4. Roh, C. J., Jung, M. C., Kim, J. R., Go, K. J., Kim, J., Oh, H. J., Jo, Y. R., Shin, Y. J., Choi, J. G., and Kim, B. J. Polar metal phase induced by oxygen octahedral network relaxation in oxide thin films. *Small* **16**, 2003055 (2020).
5. Choi, I. H., Jeong, S. G., Jeong, D. G., Seo, A., Choi, W. S., and Lee, J. S. Engineering the Coherent Phonon Transport in Polar Ferromagnetic Oxide Superlattices. *Advanced Science* **12**, 2407382 (2025).
6. Juraschek, D. M., Fechner, M., Balatsky, A. V., and Spaldin, N. A. Dynamical multiferroicity. *Physical Review Materials* **1**, 014401 (2017).
7. Kamba, S., Kempa, M., Bovtun, V., Petzelt, J., Brinkman, K., and Setter, N. Soft and central mode behaviour in $PbMg_{1/3}Nb_{2/3}O_3$ relaxor ferroelectric. *Journal of Physics: Condensed Matter* **17**, 3965 (2005).
8. Maslovskaya, A., Moroz, L., Chebotarev, A. Y., and Kovtanyuk, A. E. Theoretical and numerical analysis of the Landau–Khalatnikov model of ferroelectric hysteresis. *Communications in Nonlinear Science and Numerical Simulation* **93**, 105524 (2021).
9. Yang, F., Li, X., Talbayev, D., and Chen, L. Terahertz-induced second-harmonic generation in quantum paraelectrics: hot-phonon effect. *Physical Review Letters* **135**, 056901 (2025).
10. Shin, D., Latini, S., Schäfer, C., Sato, S. A., Baldini, E., De Giovannini, U., Hübener, H., and Rubio, A. Simulating terahertz field-induced ferroelectricity in quantum paraelectric $SrTiO_3$. *Physical Review Letters* **129**, 167401 (2022).
11. Zhang, Y., Shi, X., Sung, S. H., Li, C., Huang, H., Yu, P., and El Baggari, I. Real-space visualization of order-disorder transition in $BaTiO_3$. *Science Advances* **11**, eadx9804 (2025).
12. Kopecky, M., Fabry, J., and Kub, J. Modelling of cation displacements in $SrTiO_3$ by means of multi-energy anomalous X-ray diffuse scattering. *Journal of Applied Crystallography* **49**, 1016 (2016).
13. Uwe, H., and Sakudo, T. Stress-induced ferroelectricity and soft phonon modes in $SrTiO_3$. *Physical Review B* **13**, 271 (1976).
14. Tailliez, C., Stathopulos, A., Skupin, S., Buožius, D., Babushkin, I., Vaičaitis, V., and Bergé, L. Terahertz pulse generation by two-color laser fields with circular polarization. *New Journal of Physics* **22**, 103038 (2020).
15. Basini, M., Pancaldi, M., Wehinger, B., Udina, M., Unikandanunni, V., Tadano, T., Hoffmann, M. C., Balatsky, A. V., and Bonetti, S. Terahertz electric-field-driven dynamical multiferroicity in $SrTiO_3$. *Nature* **628**, 534 (2024).
16. Li, X., Qiu, T., Zhang, J., Baldini, E., Lu, J., Rappe, A. M., and Nelson, K. A. Terahertz field–induced ferroelectricity in quantum paraelectric $SrTiO_3$. *Science* **364**, 1079 (2019).
17. Li, X., Peng, P., Dammak, H., Geneste, G., Akbarzadeh, A., Prosandeev, S., Bellaiche, L., and Talbayev, D. Terahertz pulse induced second harmonic generation and Kerr effect in the quantum paraelectric $KTaO_3$. *Physical Review B* **107**, 064306 (2023).
18. Eremenko, V. V., and Kharchenko, N. Magneto-optics of antiferromagnets. *Physics Reports* **155**, 379 (1987).

19 M. I. Aroyo, J. M. Perez-Mato, C. Capillas, E. Kroumova, S. Ivantchev, G. Madariaga, A. Kirov, and Wondratschek, H. Bilbao Crystallographic Server I: Databases and crystallographic computing programs. *Zeitschrift fuer Kristallographie* **221**, 15 (2006).
20 Choi, I. H., Urazhdin, S., Varshney, S., Jeong, S. G., Jalan, B., and Nelson, K. A. Light-Driven Ultrafast Control of Time-Reversal Symmetry in $SrTiO_3$. *arXiv:2609.09497* (2026).